

\hyphenation{di-men-s-ion-al quad-ra-tic para-me-tri-zed 
             La-g-ran-gian La-g-ran-gians}


\magnification=\magstep1
\parskip = 1pt plus 2pt

\def\ignore#1{} \def\asis#1{#1}


\output{\ooutput}
\def\ooutput{\shipout\vbox{\makeheadline\pagebody\makefootline}%
  \advancepageno\global\fnotenumber=0
  \ifnum\outputpenalty>-20000 \else\dosupereject\fi}

\def\page{\vfill\eject}


\xdef\numcat{\catcode`0=11\catcode`1=11\catcode`2=11\catcode`3=11
\catcode`4=11
\catcode`5=11\catcode`6=11\catcode`7=11\catcode`8=11\catcode`9=11
\catcode`@=11}
\xdef\unnumcat{\catcode`0=12\catcode`1=12\catcode`2=12\catcode`3=12
\catcode`4=12
\catcode`5=12\catcode`6=12\catcode`7=12\catcode`8=12\catcode`9=12
\catcode`@=12}
{\numcat\gdef\0{0}\gdef\1{1}\gdef\2{2}\gdef\3{3}\gdef\4{4}\gdef\5{5}
\gdef\6{6}\gdef\7{7}\gdef\8{8}\gdef\9{9}}
{\catcode`@=11\expandafter\gdef\csname @0\endcsname{0}
\expandafter\gdef\csname @1\endcsname{1}\expandafter\gdef\csname
@2\endcsname{2}
\expandafter\gdef\csname @3\endcsname{3}\expandafter\gdef\csname
@4\endcsname{4}
\expandafter\gdef\csname @5\endcsname{5}\expandafter\gdef\csname
@6\endcsname{6}
\expandafter\gdef\csname @7\endcsname{7}\expandafter\gdef\csname
@8\endcsname{8}
\expandafter\gdef\csname @9\endcsname{9}}


\def\Title#1.{\def\\{\par}\vglue 30pt\title#1.\vskip20pt}

\def\Abstract#1\par{\par\bigskip\indent{\bf
Abstract.}\enspace#1\par\bigskip}

\def\shortline#1{\hbox to 3truein{#1}}
\def\shorterline#1{\hbox to 2.5truein{#1}}

\def\boxits#1{\def\\{\hfil\egroup\par\shorterline\bgroup\asis}%
\vbox{\shorterline\bgroup#1\hfil\egroup}}
\def\boxit#1{\def\\{\hfil\egroup\par\shortline\bgroup\relax
\expandafter\ignorespaces}%
\vbox{\shortline\bgroup\ignorespaces#1\hfil\egroup}}

\newcount\autnumber  \autnumber=1

\def\author#1\\#2\email#3\support#4.{%
\expandafter\def\csname autx\romannumeral\the\autnumber\endcsname
   {\boxit{#1\fnotemark\\#2\\{\tt #3}\\}
\vfootnote\fnotemark{\it Supported in part by #4.}
\global\advance\fnotenumber by 1}\global\advance\autnumber by 1}

\def\printauthor{\global\advance\autnumber by -1
\csname autform\romannumber\the\autnumber\endcsname}



\newcount\fnotenumber  \fnotenumber=0
\def\fnotemark{\ifcase\the\fnotenumber$^\dagmath$\or$^\ddagmath$\or
                        
$^\Smath$\or$^\Pmath$\or$^\|$\or$\Upstar$\fi}
\def\fnote#1{{\footnote\fnotemark{\smallstyle
#1}\global\advance\fnotenumber by 1}}

%
%

\newif\ifeqnwrite  \eqnwritefalse
\newif\ifeqnread  \eqnreadfalse
\newif\iflocal \localfalse
\newif\ifwarn \warnfalse

\def\keyout{\eqnwritetrue\immediate\openout1=\jobname.key}
\def\keyin{\numcat \eqnreadtrue \input \jobname.key \unnumcat }

\newtoks\outtoks
\def\wdef#1{\outtoks=\expandafter{\csname#1\endcsname}%
\immediate\write1{\def\the\outtoks{\csname#1\endcsname}}}
\def\wsdef#1{\outtoks=\expandafter{\csname\sectionkey#1\endcsname}%
\immediate\write1{\def\the\outtoks{\csname\sectionkey#1\endcsname}}}


\outer\def\ends{\ifeqnwrite\immediate\closeout1\fi
\vfill\supereject\end}

%
%

\newif\ifproofmode \proofmodefalse

\def\strutdepth{\dp\strutbox}
\def\margintagleft#1{\strut\vadjust{\kern-\strutdepth\vtop to
\strutdepth
   {\baselineskip\strutdepth\vss\llap{\sevenrm#1\quad}\null}}}
\def\prooflabel#1{\ifproofmode\margintagleft{#1}\fi}

\def\folio{\ifproofmode\jobname\quad\shortdate\quad\fi{\tenrm
\number\pageno}}

\def\date{\ifcase\month\or January\or February\or March\or April\or
May\or June\or July\or August\or September\or October\or November\or
December\fi
\space\number\day, \number\year}
\def\gobble#1#2{}
\def\shortdate{\number\month/\number\day/\expandafter\gobble\number
\year}

\def\dated{\footnote\ {\hfill\bigrm \date}}


\def\center{\parindent=0pt \parfillskip = 0pt 
\leftskip=0pt plus 1fil \rightskip = \leftskip \spaceskip=.3333em
\xspaceskip = .5em
\pretolerance=9999 \tolerance=9999 \hyphenpenalty=9999
\exhyphenpenalty=9999\relax}

\def\title#1.{{\center{\baselineskip=20pt \Bigrm #1}\par\nobreak}}
\def\btitle#1.{{\center{\bf #1}\par\nobreak}}
\def\ctitle#1.{{\center{\bigrm #1}\par\nobreak}}

\def\Headfont{\Bigbf} \def\headfont{\bf} \def\subheadfont{\bf}

\def\subhead#1{\medskip\heading{#1}\subheadfont}
\def\Headtagleft#1#2{\bigbreak{\parindent =
0pt\prooflabel{#2}\heading{#1}\Headfont}}
\def\headtagleft#1#2{\bigbreak{\parindent =
0pt\prooflabel{#2}\heading{#1}\headfont}}
\def\heading#1#2{{\noindent \rightskip = 0pt plus 1fil 
\spaceskip=.3333em \xspaceskip=.5em \hyphenpenalty=9999 
\exhyphenpenalty=9999 #2\hangindent=8pc
#1\par\nobreak}\medskip\noindent}

%
%
%
%
%
%
%

\newcount\chapternumber  \chapternumber=1
\newcount\sectionnumber  \sectionnumber=1
\newcount\equationnumber  \equationnumber=1
\newcount\statementnumber \statementnumber=1
\newcount\itemnumber \itemnumber=1

\newif\ifbeginningdocument \beginningdocumenttrue
\newif\iflongmode  \longmodefalse
\newif\ifbookmode \bookmodefalse

\def\chapterkey{cxyz}  
\def\sectionkey{sxyz}  

\def\chapnum{\the\chapternumber}
\def\secnum{\the\sectionnumber}
\def\eqnum{\the\equationnumber}
\def\stnum{\the\statementnumber}

\def\xcs#1{\expandafter\zcs\the#1.\aftergroup.}
\def\zcs#1{\ifx#1.\let\next=\relax\else\expandafter
\aftergroup\csname#1\endcsname\let\next=\zcs\fi\next}
\def\ncs#1{\ifx#1.\let\next=\relax\else\expandafter
\aftergroup\csname@#1\endcsname\let\next=\zcs\fi\next}

\def\cxnum{\xcs\chapternumber}
\def\scxnum{\xcs\chapternumber\xcs\sectionnumber}
\def\eqxnum{\scxnum\xcs\equationnumber}
\def\stxnum{\scxnum\xcs\statementnumber}
\def\pgnum{\the\pageno}
\def\setcsnum#1.#2.{\begingroup\aftergroup\global\aftergroup
\chapternumber
\ncs#1.\aftergroup\global\aftergroup\sectionnumber\ncs#2.\endgroup}

\def\secmode#1.#2.{\ifbookmode#1.\fi#2}
\def\sekmode#1.#2.{\ifbookmode#1.\fi\iflongmode#2.\fi}
\def\eqmode#1.#2.#3.{\ifbookmode#1.\fi\iflongmode#2.\fi#3}
\def\stmode#1 #2.#3.#4.{#1 \ifbookmode#2.\fi\iflongmode#3.\fi#4}
\def\stnmode#1 #2.#3.#4.{\ifbookmode#2.\fi\iflongmode#3.\fi#4}

\def\cinfo#1.#2.#3.{\expandafter\xdef\csname c@#1\endcsname{#2}%
\expandafter\xdef\csname s@#1\endcsname{1}%
\equationnumber=1\statementnumber=1\pageno=#3}
\def\sinfo#1.#2.#3.{\expandafter\xdef\csname s@#1\endcsname{#2}%
\equationnumber=1\statementnumber=1\pageno=#3}
\def\zinfo#1.#2.#3.#4.#5.{\expandafter\xdef\csname c@#1\endcsname{#2}%
\expandafter\xdef\csname s@#3\endcsname{#4}%
\equationnumber=1\statementnumber=1\pageno=#5}
\def\ainfo#1.#2.#3.#4.#5.#6.#7.{\expandafter\xdef\csname
c@#1\endcsname{#2}%
\expandafter\xdef\csname s@#3\endcsname{#4}%
\equationnumber=#5\statementnumber=#6\pageno=#7}

\def\Chapter#1#2.{{\numcat\gdef\chapterkey{#1}\ifbeginningdocument
   \else\global\advance\chapternumber by 1\fi
\global\beginningdocumentfalse
\begingroup\aftergroup\xdef\aftergroup\nuxxx\aftergroup{
\cxnum\aftergroup}\endgroup
\expandafter\ifx\csname c@#1\endcsname\relax\else
\edef\eqxxx{\csname c@#1\endcsname}%
\ifx\eqxxx\nuxxx\else\immediate\write0{Warning: Chapter #1 already
exists}\fi\fi
\global\sectionnumber=0\global\equationnumber=1\global
\statementnumber=1
\gdef\sectionkey{#1}%
\expandafter\xdef\csname c@#1\endcsname{\nuxxx}
\ifeqnwrite\wdef{c@#1}\wdef{s@#1}\fi  
\immediate\write0{***  Chapter \chapterkey\space ***}%
\Headtagleft{Chapter\kern.4em\the\chapternumber. #2.}{#1}}}

\def\NChapter#1#2#3.{{\numcat\gdef\chapterkey{#2}\global
\chapternumber=#1\global\beginningdocumentfalse
\begingroup\aftergroup\xdef\aftergroup\nuxxx\aftergroup{
\cxnum\aftergroup}\endgroup
\expandafter\ifx\csname c@#2\endcsname\relax\else
\edef\eqxxx{\csname c@#2\endcsname}%
\ifx\eqxxx\nuxxx\else\immediate\write0{Warning: Chapter #2 already
exists}\fi\fi
\global\sectionnumber=0\global\equationnumber=1\global
\statementnumber=1\gdef\sectionkey{#2}%
\expandafter\xdef\csname c@#2\endcsname{\nuxxx}%
\ifeqnwrite\wdef{c@#2}\wdef{s@#2}\fi  
\immediate\write0{***  Chapter \chapterkey\space ***}%
\Headtagleft{Chapter\kern.4em\the\chapternumber. #3.}{#2}}}

\def\chapter#1{{\numcat Chapter \expandafter\ifx\csname
c@#1\endcsname\relax
  \immediate\write0{Warning: Chapter #1 not defined}{\bf #1}\else
  \csname c@#1chapter\fi}}

\def\Section#1#2. {{\numcat\gdef\sectionkey{#1}\ifbeginningdocument
   \else\global\advance\sectionnumber by 1\fi
\global\beginningdocumentfalse
\begingroup\aftergroup\xdef\aftergroup\nuxxx\aftergroup{%
\scxnum\aftergroup}\endgroup
\expandafter\ifx\csname s@#1\endcsname\relax\else
\edef\eqxxx{\csname s@#1\endcsname}%
\ifx\eqxxx\nuxxx\else\immediate\write0{Warning: Section #1 already
exists}\fi\fi
\iflongmode\global\equationnumber=1\global\statementnumber=1\fi
\expandafter\xdef\csname s@#1\endcsname{\nuxxx}
\ifeqnwrite\wdef{s@#1}\fi  
\immediate\write0{***  Section \sectionkey\space ***}%
\headtagleft{\expandafter\secmode\nuxxx. #2}{#1}}}

\def\NSection#1#2#3.{{\numcat\gdef\sectionkey{#2}\global
\sectionnumber=#1
\global\beginningdocumentfalse
\begingroup\aftergroup\xdef\aftergroup\nuxxx\aftergroup{
\scxnum\aftergroup}\endgroup
\expandafter\ifx\csname s@#2\endcsname\relax\else
\edef\eqxxx{\csname s@#2\endcsname}%
\ifx\eqxxx\nuxxx\else\immediate\write0{Warning: Section #2 already
exists}\fi\fi
\iflongmode\global\equationnumber=1\global\statementnumber=1\fi
\expandafter\xdef\csname s@#2\endcsname{\nuxxx}
\ifeqnwrite\wdef{s@#2}\fi  
\immediate\write0{***  Section \sectionkey\space ***}%
\headtagleft{\expandafter\secmode\nuxxx. #3.}{#2}}}

\def\section#1{{\numcat Section \expandafter\ifx\csname
s@#1\endcsname\relax
  \immediate\write0{Warning: Section #1 not defined}{\bf #1}\else
  \edef\nuxxx{\csname s@#1\endcsname}\expandafter\secmode\nuxxx\fi}}

\def\Subsection#1.{\subhead{#1}}

\def\addkeyz#1.#2\addkeyz{\ifx.#2.\sectionkey.#1.\else#1.#2\fi}

%
%

\def\Eq{\eqtag}
\def\eqtag#1{\newequation\eqno{#1}()}
\def\nqtag#1{\newequation\zlap{#1}()}

\def\Eql#1#2$${\Eq{#1}\,\kern -\displaywidth
     \rlap{\quad $\displaystyle#2$}\kern \displaywidth \kern -1pt$$}
\def\El#1$${\eqno\,\kern -\displaywidth
     \rlap{\quad $\displaystyle#1$}\kern \displaywidth \kern -1pt$$}

\def\zlap#1{\llap{#1\kern-3pt}}

\def\newequation#1#2#3#4{\prooflabel{#2}{\numcat
\begingroup\aftergroup\xdef\aftergroup\nuxxx\aftergroup{
\eqxnum\aftergroup}\endgroup
\expandafter\ifx\csname e@#2\endcsname\relax\else
\edef\eqxxx{\csname e@#2\endcsname}%
\ifx\eqxxx\nuxxx\else\immediate\write0{Warning: Equation #2 already
exists}\fi\fi
\expandafter\xdef\csname e@#2\endcsname{\nuxxx}
\ifeqnwrite\wdef{e@#2}\fi}%
#1{#3\expandafter\eqmode\nuxxx#4}\global\advance\equationnumber by 1}

\def\eq#1{\eqcite{#1}()}

\def\eqcite#1#2#3{{\numcat\expandafter\ifx\csname e@#1\endcsname
\relax
\immediate\write0{Warning: Equation #1 not
defined.}\edef\eqxxx{#2#1\blackbox#3}\else
\edef\nuxxx{\csname e@#1\endcsname}%
\edef\eqxxx{#2\expandafter\eqmode\nuxxx#3}\fi{\rm\eqxxx}}}

%
%
%

\def\addtab#1={#1\;&=}

\def\aeq#1{\def\\{\cr}\eqalign{#1\cr}}

\def\eeqn#1{\def\\{\cr\addtab}\def\Eq{&\nqtag}\openup1\jot
\tabskip=\centering 
  \halign to \displaywidth{$\hfil\displaystyle
   ##$\tabskip=0pt&$\displaystyle##\hfil$\tabskip=\centering
   &##\tabskip=0pt\cr\addtab#1\cr}\def\Eq{\eqtag}}

\def\Eeqn#1{\def\\{\cr}\def\Eq{&\nqtag}\openup1\jot
\tabskip=\centering 
  \halign to \displaywidth{$\hfil\displaystyle
   ##$\tabskip=0pt&$\displaystyle##\hfil$\tabskip=\centering
   &##\tabskip=0pt\cr#1\cr}\def\Eq{\eqtag}}

\def\ntable#1{\itemnumber=1 
   \def\\{\cr \the\itemnumber.\global\advance\itemnumber by 1
&}\vcenter{\openup1\jot 
   \halign{##&&\quad $\displaystyle{}##\hfil$\cr 
   \the\itemnumber.\global\advance\itemnumber by 1 &#1\cr}}}

\def\items#1{\itemnumber=1 
   \def\\{\par
\item{\ro(\the\itemnumber\ro)}\enspace\advance\itemnumber by
1}\\#1\par}
\def\ritems#1{\itemnumber=1 \def\\{\par \item{\ro (\sl  
   \romannumber{\the\itemnumber}\/\ro )}\enspace\advance\itemnumber
by 1}\\#1\par}
\def\aitems#1{\itemnumber=1  \def\\{\par \item{\ro (\sl 
  \alphanumber{\the\itemnumber}\/\ro )}\enspace\advance\itemnumber by
1}\\#1\par}

%
%
%

\def\declare#1#2{\medbreak\noindent{\bf#1.\kern.2em}#2\par\medbreak}
\def\exdeclare#1#2#3\par{\declare{\Statement{#1}{#2}}{#3}}
\def\stdeclare#1#2#3\par{\exdeclare{#1}{#2}{{\sl #3}}\par}

\def\Statement#1#2{\prooflabel{#2}{\numcat
\begingroup\aftergroup\xdef\aftergroup\nuxxx\aftergroup{%
\stxnum\aftergroup}\endgroup \edef\stxxx{#1 \nuxxx}%
\expandafter\ifx\csname t@#2\endcsname\relax\else
\edef\eqxxx{\csname t@#2\endcsname}%
\ifx\stxxx\eqxxx\else\immediate\write0{Warning: Statement #2 already
exists}\fi\fi
\expandafter\xdef\csname t@#2\endcsname{\stxxx}%
\ifeqnwrite\wdef{t@#2}\fi}%
#1\kern.4em\expandafter\eqmode\nuxxx\global\advance\statementnumber
by 1}

\def\stwrite#1#2{\stlabel{#1}{#2}\stmode}

\def\stlabel#1#2#3{{\numcat\expandafter\ifx\csname
t@#2\endcsname\relax
\immediate\write0{Warning: #1 #2 not defined.}\edef\stxxx{#1
#2\blackbox}\else
\edef\nuxxx{\csname t@#2\endcsname}%
\edef\stxxx{\expandafter#3\nuxxx}\fi{\stxxx}}}

\def\Lm#1{\stdeclare{Lemma}{#1}} \def\lm{\stwrite{Lemma}}
\def\Lemma{\Lm} 

\def\Pr#1{\stdeclare{Proposition}{#1}} \def\pr{\stwrite{Proposition}}

\def\Co#1{\stdeclare{Corollary}{#1}} \def\co{\stwrite{Corollary}}

\def\Df#1{\exdeclare{Definition}{#1}} \def\df{\stwrite{Definition}}

\def\Remark #1\par{\medbreak\noindent{\it
Remark\/}:\enspace{#1}\par\medbreak}
\def\Warning #1\par{\medbreak\noindent{\it
Warning\/}:\enspace{#1}\par\medbreak}

\def\Proof{\medbreak\indent{\it Proof\/}:\enspace}
\def\qed{\hfill\parfillskip=0pt {\it Q.E.D.}\par\parfillskip=0pt plus 1fil\medbreak}

%
%
%
%

\newcount\refnumber \refnumber=1
\newif\ifrefmode \newif\ifrefs

\def\refsin{{\refmodefalse \input\jobname.ref }\refstrue}

\def\refstyle{\parskip=1pt \frenchspacing
\rightskip=0pt plus 4em \spaceskip = .3333em \xspaceskip = .5em
\pretolerance=9999 \tolerance=9999
\hyphenpenalty=9999 \exhyphenpenalty=9999
\interlinepenalty=1000 \raggedbottom}

\def\References{{\refmodetrue \title References. \vskip 15pt
{\refstyle \input \jobname.ref }}}

\def\book#1;#2;#3\par{\ifrefmode\namez#1;{\it #2},#3.\par\fi}
\def\paper#1;#2;#3; #4(#5)#6\par{\ifrefmode\namez#1;#2,{\it #3}\if
a#4, to appear.\else
     { \bf #4}(#5),#6.\fi\par\fi}
\def\preprint#1;#2;#3\par{\ifrefmode\namez#1;#2, preprint,#3.\par\fi}
\def\inbook#1;#2;#3;#4\par{\ifrefmode\namez#1;#2, {\sl in\/}:{\it
#3},#4.\par\fi}
\def\appendix#1;#2;#3;#4\par{\ifrefmode\namez#1;#2, {\sl appendix
in\/}:{\it #3},#4.\par \fi}
\def\other#1\par{\ifrefmode \hangindent\parindent #1.\par\fi}

\def\key#1 {\ifrefmode \hangindent2\parindent \textindent
{\ifrefs \ifproofmode{\fiverm #1} \fi\fi  \rf{#1}}\else
{\numcat\expandafter\xdef\csname refx#1\endcsname{\the\refnumber}%
\global\advance\refnumber by 1\ifeqnwrite\wdef{refx#1}\fi}\fi}

\newif\ifnames \namesfalse
\def\namez#1;{\namesfalse\namezz#1,;} 
\def\namezz#1,#2,#3;{\ifx,#3,\ifnames \ and\fi\fi #1,#2,\ifx,#3,\else 
    \namestrue\namezz #3;\fi}

\def\rf#1{\rfz#1;\rfz}
\def\rfz#1;#2\rfz{{\numcat\def\rfxxx{#1}\ifrefs\expandafter\ifx
   \csname refx#1\endcsname\relax\immediate\write0{Reference  [#1] 
not provided.}\else 
   \xdef\rfxxx{\csname refx#1\endcsname}\fi \fi[{\bf
\rfxxx}\ifx;#2;\else ;\nosemiz#2\fi]}}
\def\nosemiz#1;{#1}

\newread\epsffilein    
\newif\ifepsffileok    
\newif\ifepsfbbfound   
\newif\ifepsfverbose   
\newdimen\epsfxsize    
\newdimen\epsfysize    
\newdimen\epsftsize    
\newdimen\epsfrsize    
\newdimen\epsftmp      
\newdimen\pspoints     
\pspoints=1bp          
\epsfxsize=0pt         
\epsfysize=0pt         
\def\epsfbox#1{\global\def\epsfllx{72}\global\def\epsflly{72}%
   \global\def\epsfurx{540}\global\def\epsfury{720}%
   \def\lbracket{[}\def\testit{#1}\ifx\testit\lbracket
   \let\next=\epsfgetlitbb\else\let\next=\epsfnormal\fi\next{#1}}%
\def\epsfgetlitbb#1#2 #3 #4 #5]#6{\epsfgrab #2 #3 #4 #5 .\\%
   \epsfsetgraph{#6}}%
\def\epsfnormal#1{\epsfgetbb{#1}\epsfsetgraph{#1}}%
\def\epsfgetbb#1{%
%
%
\openin\epsffilein=#1
\ifeof\epsffilein\errmessage{I couldn't open #1, will ignore it}\else
%
%
   {\epsffileoktrue \chardef\other=12
    \def\do##1{\catcode`##1=\other}\dospecials \catcode`\ =10
    \loop
       \read\epsffilein to \epsffileline
       \ifeof\epsffilein\epsffileokfalse\else
%
%
          \expandafter\epsfaux\epsffileline:. \\%
       \fi
   \ifepsffileok\repeat
   \ifepsfbbfound\else
    \ifepsfverbose\message{No bounding box comment in #1; using
defaults}\fi\fi
   }\closein\epsffilein\fi}%
%
%
\def\epsfsetgraph#1{%
   \epsfrsize=\epsfury\pspoints
   \advance\epsfrsize by-\epsflly\pspoints
   \epsftsize=\epsfurx\pspoints
   \advance\epsftsize by-\epsfllx\pspoints
%
%
   \epsfsize\epsftsize\epsfrsize
   \ifnum\epsfxsize=0 \ifnum\epsfysize=0
      \epsfxsize=\epsftsize \epsfysize=\epsfrsize
%
%
     \else\epsftmp=\epsftsize \divide\epsftmp\epsfrsize
       \epsfxsize=\epsfysize \multiply\epsfxsize\epsftmp
       \multiply\epsftmp\epsfrsize \advance\epsftsize-\epsftmp
       \epsftmp=\epsfysize
       \loop \advance\epsftsize\epsftsize \divide\epsftmp 2
       \ifnum\epsftmp>0
          \ifnum\epsftsize<\epsfrsize\else
             \advance\epsftsize-\epsfrsize \advance\epsfxsize\epsftmp \fi
       \repeat
     \fi
   \else\epsftmp=\epsfrsize \divide\epsftmp\epsftsize
     \epsfysize=\epsfxsize \multiply\epsfysize\epsftmp   
     \multiply\epsftmp\epsftsize \advance\epsfrsize-\epsftmp
     \epsftmp=\epsfxsize
     \loop \advance\epsfrsize\epsfrsize \divide\epsftmp 2
     \ifnum\epsftmp>0
        \ifnum\epsfrsize<\epsftsize\else
           \advance\epsfrsize-\epsftsize \advance\epsfysize\epsftmp
           \fi
     \repeat     
   \fi
%
%
   \ifepsfverbose\message{#1: width=\the\epsfxsize,
    height=\the\epsfysize}\fi
   \epsftmp=10\epsfxsize \divide\epsftmp\pspoints
   \vbox to\epsfysize{\vfil\hbox to\epsfxsize{%
      \special{illustration #1 scaled \number\epsfscale}
      \hfil}}%
\epsfxsize=0pt\epsfysize=0pt\epsfscale=1000 }%

%
%
{\catcode`\%=12
\global\let\epsfpercent=
%
%
\long\def\epsfaux#1#2:#3\\{\ifx#1\epsfpercent
   \def\testit{#2}\ifx\testit\epsfbblit
      \epsfgrab #3 . . . \\%
      \epsffileokfalse
      \global\epsfbbfoundtrue
   \fi\else\ifx#1\par\else\epsffileokfalse\fi\fi}%
%
%
\def\epsfgrab #1 #2 #3 #4 #5\\{%
   \global\def\epsfllx{#1}\ifx\epsfllx\empty
      \epsfgrab #2 #3 #4 #5 .\\\else
   \global\def\epsflly{#2}%
   \global\def\epsfurx{#3}\global\def\epsfury{#4}\fi}%
%
%
%
%

\newcount\epsfscale    
\newdimen\epsftmpp     
\newdimen\epsftmppp    
\newdimen\epsfM        
\newdimen\sppoints     
\epsfscale=1000        
\sppoints=1000sp       
\epsfM=1000\sppoints
%
\def\computescale#1#2{%
  \epsftmpp=#1 \epsftmppp=#2
  \epsftmp=\epsftmpp \divide\epsftmp\epsftmppp  
  \epsfscale=\epsfM \multiply\epsfscale\epsftmp 
  \multiply\epsftmp\epsftmppp                   
  \advance\epsftmpp-\epsftmp                    
  \epsftmp=\epsfM                               
  \loop \advance\epsftmpp\epsftmpp              
    \divide\epsftmp 2                           
    \ifnum\epsftmp>0
      \ifnum\epsftmpp<\epsftmppp\else           
        \advance\epsftmpp-\epsftmppp            
        \advance\epsfscale\epsftmp \fi          
  \repeat
  \divide\epsfscale\sppoints}
\def\epsfsize#1#2{%
  \ifnum\epsfscale=1000
    \ifnum\epsfxsize=0
      \ifnum\epsfysize=0
      \else \computescale{\epsfysize}{#2}
      \fi
    \else \computescale{\epsfxsize}{#1}
    \fi
  \else
    \epsfxsize=#1
    \divide\epsfxsize by 1000 \multiply\epsfxsize by \epsfscale
  \fi}


\epsfscale = 833  


\newcount\figurenumber \figurenumber=1
\def\fgnum{\the\figurenumber}

\def\Fig#1#2{\prooflabel{#1}\edef\nuxxx{\fgnum}%
\expandafter\ifx\csname fg#1\endcsname\relax\else
\edef\eqxxx{\csname fg#1\endcsname}%
\ifx\nuxxx\eqxxx\else\immediate\write0{Figure #1 already exists}\fi
\fi
\expandafter\xdef\csname fg#1\endcsname{\nuxxx}%
\insert\topins{\penalty100 \splittopskip=0pt \splitmaxdepth=\maxdimen
\floatingpenalty=0
\hfil \epsfbox{#1.ps}\hfill \par\nobreak \vskip10pt%
\center{\bigrm Figure \nuxxx. \quad #2.}\par\vskip15pt}%
\ifeqnwrite\wsdef{fg#1}\fi\global\advance\figurenumber by 1}

\def\fig#1{\expandafter\ifx\csname fg#1\endcsname\relax
\iflocal\def\bbxxx{}\else
\immediate\write0{Figure #1 not defined.}\def\bbxxx{\blackbox}\fi
\edef\nuxxx{\bbxxx #1}\else\edef\nuxxx{\csname fg#1\endcsname}\fi
Figure \nuxxx}

%
%
%

\newif\ifindex \indexfalse
\def\indexout{\indextrue\immediate\openout2=\jobname.ind%
\openout3=\jobname.aut\input \jobname.rfa}


\def\hexnumber#1{\ifcase#1 0\or1\or2\or3\or4\or5\or6\or7\or8\or9\or
 A\or B\or C\or D\or E\or F\fi}
\def\romannumber#1{\romannumeral#1}
\def\alphanumber#1{\count255 = #1 \advance\count255 by 96
\expandafter\char\count255}


\def\ro#1{\hbox{\rm #1}}

\def\bo#1{{\bf #1}}

\font\bigrm=cmr10 scaled\magstep1
\font\Bigrm=cmr10 scaled\magstep2

\font\Bigbf=cmbx10 scaled\magstep2
\font\elevenrm=cmr10 at 11pt

\font\ninerm=cmr9
\font\ninei=cmmi9
\font\ninesy=cmsy9
\font\ninebf=cmbx9
\font\ninett=cmtt9
\font\ninesl=cmsl9
\font\nineit=cmti9

\font\nineex=cmex10 at 9pt



\fontdimen16\tensy=2.7pt \fontdimen17\tensy=2.7pt 

\def\sup#1{\raise 2.7pt \hbox{\sevenrm #1}}
\def\sub#1{\lower 2.7pt \hbox{\sevenrm #1}}


\font\tenib=cmmib10 \font\sevenib=cmmib10 at 7pt \font\fiveib=cmmib10
at 5pt
\newfam\ibfam
\textfont\ibfam=\tenib  \scriptfont\ibfam=\sevenib
\scriptscriptfont\ibfam=\fiveib

\edef\ibhx{\hexnumber\ibfam}

\mathchardef\alphaB="0\ibhx0B
\mathchardef\betaB="0\ibhx0C
\mathchardef\gammaB="0\ibhx0D
\mathchardef\deltaB="0\ibhx0E
\mathchardef\epsilonB="0\ibhx0F
\mathchardef\zetaB="0\ibhx10
\mathchardef\etaB="0\ibhx11
\mathchardef\thetaB="0\ibhx12
\mathchardef\iotaB="0\ibhx13
\mathchardef\kappaB="0\ibhx14
\mathchardef\lambdaB="0\ibhx15
\mathchardef\muB="0\ibhx16
\mathchardef\nuB="0\ibhx17
\mathchardef\xiB="0\ibhx18
\mathchardef\piB="0\ibhx19
\mathchardef\rhoB="0\ibhx1A
\mathchardef\sigmaB="0\ibhx1B
\mathchardef\tauB="0\ibhx1C
\mathchardef\upsilonB="0\ibhx1D
\mathchardef\phiB="0\ibhx1E
\mathchardef\chiB="0\ibhx1F
\mathchardef\psiB="0\ibhx20
\mathchardef\omegaB="0\ibhx21
\mathchardef\varepsilonB="0\ibhx22
\mathchardef\varthetaB="0\ibhx23
\mathchardef\varpiB="0\ibhx24
\mathchardef\varrhoB="0\ibhx25
\mathchardef\varsigmaB="0\ibhx26
\mathchardef\varphiB="0\ibhx27

\mathchardef\GammaBI="0\ibhx00
\mathchardef\DeltaBI="0\ibhx01
\mathchardef\ThetaBI="0\ibhx02
\mathchardef\LambdaBI="0\ibhx03
\mathchardef\XiBI="0\ibhx04
\mathchardef\PiBI="0\ibhx05
\mathchardef\SigmaBI="0\ibhx06
\mathchardef\UpsilonBI="0\ibhx07
\mathchardef\PhiBI="0\ibhx08
\mathchardef\PsiBI="0\ibhx09
\mathchardef\OmegaBI="0\ibhx0A


\font\teneuf=eufm10 \font\seveneuf=eufm7 \font\fiveeuf=eufm5
\newfam\euffam
\textfont\euffam=\teneuf \scriptfont\euffam=\seveneuf
\scriptscriptfont\euffam=\fiveeuf

\def\frak#1{{\fam\euffam\relax#1}}


\def\Cal#1{{\cal#1}}

\def\CA{\Cal A}


\font\tenmsb=msbm10 \font\sevenmsb=msbm7 \font\fivemsb=msbm5
\newfam\msbfam  \textfont\msbfam=\tenmsb  
   \scriptfont\msbfam=\sevenmsb \scriptscriptfont\msbfam=\fivemsb

\edef\msbhx{\hexnumber\msbfam}
\def\Bbb#1{{\fam\msbfam\relax#1}}

\def\R{\Bbb R}

\def\Id{\hbox{\tenrm 1\kern-3.8pt \elevenrm1}}


\def\Gbar{\kern.1em \overline{\kern-.1em G}{}}

\def\Htilde{\widetilde
H{}}

\def\Vtilde{\widetilde V{}}

\def\hats{
\def\ahat{\hat a{}}\def\bhat{\hat b{}}\def\chat{\hat c{}}
\def\dhat{\hat d{}}\def\ehat{\hat e{}}\def\fhat{\hat f{}}
\def\ghat{\hat g{}}\def\hhat{\hat h{}}
\def\ihat{\hat \imath{}}\def\jhat{\hat \jmath{}}
\def\khat{\hat k{}}\def\lhat{\hat l{}}
\def\mhat{\hat m{}}\def\nhat{\hat n{}}\def\ohat{\hat o{}}
\def\phat{\hat p{}}\def\qhat{\hat q{}}\def\rhat{\hat r{}}
\def\shat{\hat s{}}\def\that{\hat t{}}\def\uhat{\hat u{}}
\def\vhat{\hat v{}}\def\what{\hat w{}}
\def\xhat{\hat x{}}\def\yhat{\hat y{}}\def\zhat{\hat z{}}}

\def\widehats{
\def\ahat{\widehat a{}}\def\bhat{\widehat b{}}\def\chat{\widehat c{}}
\def\dhat{\widehat d{}}\def\ehat{\widehat e{}}\def\fhat{\widehat f{}}
\def\ghat{\widehat g{}}\def\hhat{\widehat h{}}
\def\ihat{\widehat \imath{}}\def\jhat{\widehat \jmath{}}
\def\khat{\widehat k{}}\def\lhat{\widehat l{}}
\def\mhat{\widehat m{}}\def\nhat{\widehat n{}}\def\ohat{\widehat o{}}
\def\phat{\widehat p{}}\def\qhat{\widehat q{}}\def\rhat{\widehat r{}}
\def\shat{\widehat s{}}\def\that{\widehat t{}}\def\uhat{\widehat u{}}
\def\vhat{\widehat v{}}\def\what{\widehat w{}}
\def\xhat{\widehat x{}}\def\yhat{\widehat y{}}\def\zhat{\widehat z{}}}

\def\tildes{
\def\atilde{\tilde a{}}\def\btilde{\tilde b{}}\def\ctilde{\tilde c{}}
\def\dtilde{\tilde d{}}\def\etilde{\tilde e{}}\def\ftilde{\tilde f{}}
\def\gtilde{\tilde g{}}\def\htilde{\tilde h{}}
\def\itilde{\tilde \imath{}}\def\jtilde{\tilde \jmath{}}
\def\ktilde{\tilde k{}}\def\ltilde{\tilde l{}}
\def\mtilde{\tilde m{}}\def\ntilde{\tilde n{}}\def\otilde{\tilde o{}}
\def\ptilde{\tilde p{}}\def\qtilde{\tilde q{}}\def\rtilde{\tilde r{}}
\def\stilde{\tilde s{}}\def\ttilde{\tilde t{}}\def\utilde{\tilde u{}}
\def\vtilde{\tilde v{}}\def\wtilde{\tilde w{}}
\def\xtilde{\tilde x{}}\def\ytilde{\tilde y{}}
\def\ztilde{\tilde z{}}}

\def\widetildes{
\def\atilde{\widetilde a{}}\def\btilde{\widetilde
b{}}\def\ctilde{\widetilde c{}}
\def\dtilde{\widetilde d{}}\def\etilde{\widetilde
e{}}\def\ftilde{\widetilde f{}}
\def\gtilde{\widetilde g{}}\def\htilde{\widetilde h{}}
\def\itilde{\widetilde \imath{}}\def\jtilde{\widetilde \jmath{}}
\def\ktilde{\widetilde k{}}\def\ltilde{\widetilde l{}}
\def\mtilde{\widetilde m{}}\def\ntilde{\widetilde
n{}}\def\otilde{\widetilde o{}}
\def\ptilde{\widetilde p{}}\def\qtilde{\widetilde
q{}}\def\rtilde{\widetilde r{}}
\def\stilde{\widetilde s{}}\def\ttilde{\widetilde
t{}}\def\utilde{\widetilde u{}}
\def\vtilde{\widetilde v{}}\def\wtilde{\widetilde w{}}
\def\xtilde{\widetilde x{}}\def\ytilde{\widetilde
y{}}\def\ztilde{\widetilde z{}}}

\def\bars{
\def\abar{\bar a{}}\def\bbar{\bar b{}}\def\cbar{\bar c{}}
\def\dbar{\bar d{}}\def\ebar{\bar e{}}\def\fbar{\bar f{}}
\def\gbar{\bar g{}}\def\hbar{\bar h{}}
\def\ibar{\bar \imath{}}\def\jbar{\bar \jmath{}}
\def\kbar{\bar k{}}\def\lbar{\bar l{}}
\def\mbar{\bar m{}}\def\nbar{\bar n{}}\def\obar{\bar o{}}
\def\pbar{\bar p{}}\def\qbar{\bar q{}}\def\rbar{\bar r{}}
\def\sbar{\bar s{}}\def\tbar{\bar t{}}\def\ubar{\bar u{}}
\def\vbar{\bar v{}}\def\wbar{\barr w{}}
\def\xbar{\bar x{}}\def\ybar{\bar y{}}\def\zbar{\bar z{}}}

\def\barr#1{\overline{#1}{}}

\def\barrs{
\def\abar{\barr a{}}\def\bbar{\barr b{}}\def\cbar{\barr c{}}
\def\dbar{\barr d{}}\def\ebar{\barr e{}}\def\fbar{\barr f{}}
\def\gbar{\barr g{}}\def\hbar{\barr h{}}
\def\ibar{\barr \imath{}}\def\jbar{\barr \jmath{}}
\def\kbar{\barr k{}}\def\lbar{\barr l{}}
\def\mbar{\barr m{}}\def\nbar{\barr n{}}\def\obar{\barr o{}}
\def\pbar{\barr p{}}\def\qbar{\barr q{}}\def\rbar{\barr r{}}
\def\sbar{\barr s{}}\def\tbar{\barr t{}}\def\ubar{\barr u{}}
\def\vbar{\barr v{}}\def\wbar{\barrr w{}}
\def\xbar{\barr x{}}\def\ybar{\barr y{}}\def\zbar{\barr z{}}}


\tildes
\hats
\bars


\font\eightex=cmex10 scaled 800 \font\sixex=cmex10 scaled 600
  \font\fourex=cmex10 scaled 400
\newfam\exsfam
\textfont\exsfam=\eightex \scriptfont\exsfam=\sixex
\scriptscriptfont\exsfam=\fourex
\edef\exshx{\hexnumber\exsfam}

\mathchardef\odotsy="220C
\mathchardef\bigvee="1\exshx57
\mathchardef\bigwedge="1\exshx56
\mathchardef\bigcap="1\exshx54
\mathchardef\bigcup="1\exshx53
\mathchardef\bigotimes="1\exshx4E
\mathchardef\bigodot="1\exshx4A
\mathchardef\Bigotimes="1\exshx4F
\mathchardef\Bigoplus="1\exshx4D
\mathchardef\Bigodot="1\exshx4B
\mathchardef\Bigwedge="1\exshx5E

\def\odot{\raise 1pt \hbox{$\,\scriptscriptstyle\odotsy\,$}}
\def\tensor{\raise 1pt \hbox{$\,\scriptscriptstyle\otimes\,$}}
\def\Wedge{\raise 1pt \hbox{$\bigwedge$}}
\def\Odot{\raise 1pt \hbox{$\bigodot$}}
\def\Tensor{\raise 1pt \hbox{$\bigotimes$}}

\def\comp{\raise 1pt \hbox{$\,\scriptstyle\circ\,$}}

\def\Upstar{^{\displaystyle *}}

\mathchardef\semidirect="2\msbhx6E
\mathchardef\directsemi="2\msbhx6F
\mathchardef\emptyset="0\msbhx3F
\mathchardef\subsetneq="2\msbhx28
\mathchardef\dagmath="0279\mathchardef\ddagmath="027A
\mathchardef\Smath="0278
\mathchardef\Pmath="027B 

\def\interior{\mathbin{\hbox{\hbox{{\vbox
    {\hrule height.4pt width6pt depth0pt}}}\vrule
    height6pt width.4pt depth0pt}\,}}

\def\blackbox{\vrule height7pt width5pt depth0pt}


\def\smallstyle{\textfont0=\ninerm \scriptfont0=\sevenrm
\scriptscriptfont0=\fiverm
\def\rm{\fam0\ninerm}\textfont1=\ninei \scriptfont1=\seveni
\scriptscriptfont1=\fivei
\textfont2=\ninesy \scriptfont2=\sevensy \scriptscriptfont2=\fivesy
\textfont3=\nineex \scriptfont3=\nineex \scriptscriptfont3=\nineex
\def\it{\fam\itfam\nineit}\textfont\itfam=\nineit
\def\sl{\fam\slfam\ninesl}\textfont\slfam=\ninesl
\def\bf{\fam\bffam\ninebf}\textfont\bffam=\ninebf
\scriptfont\bffam=\sevenbf
\scriptscriptfont\bffam=\fivebf
\def\tt{\fam\ttfam\ninett}\textfont\ttfam=\ninett
\lineskip=0pt\baselineskip=9pt\parskip=0pt\ninerm}


\def\frac#1#2{{#1\over #2}}

\def\supfr#1#2{\raise 6pt \hbox{$\scriptstyle #1\above.1pt
     \raise .4pt\hbox{$\scriptstyle #2$}$}}
\def\supp#1{\raise 7pt \hbox{$\scriptstyle #1$}}

\def\array#1{\null\,\vcenter{\normalbaselines\mathsurround=0pt
   
\ialign{$##$\hfil&&\quad$##$\hfil\crcr\mathstrut\crcr
\noalign{\kern-\baselineskip}
      #1\crcr\mathstrut\crcr\noalign{\kern-\baselineskip}}}\,}
\def\Array#1{\null\,\vcenter{\normalbaselines\mathsurround=0pt
   
\ialign{$##$\hfil&&\qquad$##$\hfil\crcr\mathstrut\crcr
\noalign{\kern-\baselineskip}
      #1\crcr\mathstrut\crcr\noalign{\kern-\baselineskip}}}\,}


\def\odow#1{\mathchoice{d^2 \over d#1^2}%
{d^2 /d#1^2}{d^2 /d#1^2}{d^2 /d#1^2}}
\def\odot{\odow}

\def\d{\, d}


\def\operator#1{\expandafter\def\csname#1\endcsname
{\mathop{\ro{#1}}\nolimits}}

\operator{order} \operator{Order} \operator{ord}
\operator{deg} \operator{Degree}
\operator{diag}
\operator{rank} \operator{det}
\operator{im}  \operator{Re} \operator{Im}
\operator{Span}
\operator{sign} \operator{wt}
\operator{Div} \operator{div} \operator{grad}
\operator{Ad}
\operator{tr}
\operator{Hom} \operator{Ker}
\operator{mod}
\operator{sech} \operator{csch}
\operator{arcsinh} \operator{arccosh}
\operator{Im}\operator{Re}


 \def\gl#1{\frak{gl}(#1)}

  \def\sL#1{\frak{sl}(#1)}


\def\all{\roq{for all}}

\def\spaper#1;#2; #3(#4)#5\par{\ifrefmode\namez#1;#2\if a#3, to
appear.\else
     { \bf #3}(#4),#5.\fi\par\fi}%

5/9/95 modifications:

\def\Section#1#2. {...
\headtagleft{\expandafter\secmode\nuxxx. #2}{#1}}}
\def\heading#1#2{...\noindent}
\def\declare#1#2{\medbreak\noindent...}
\def\Remark #1\par{\medbreak\noindent...}
\def\Warning #1\par{\medbreak\noindent...}

\let\isloaded{}
Added 

5/15/95
\spaper

1/11/96

\def\Eeqn{...} 

1/17/96

Commented out definitions of \oplus, \Oplus, \oplussy, \bigoplus

2/20/96 Change \forall to \all, \def\all{\roqq{for all}}

3/15/96 \def\other#1\par{\ifrefmode \hangindent\parindent #1.\par\fi}

10/6/96 Modified def. of \author (Supported in Part -> Supported in
\part)
\overfullrule=0pt
\hyphenation{Mou-tard}
\def\inner#1#2{\left\langle#1,#2\right\rangle}
\let\Ht\Htilde \def\Hti{\Ht^{(1)}} \def\Htii{\Ht^{(2)}}
\def\hti{\Ht^{(i)}}
\def\pt{\tilde\psi}
\def\Hi{H^{(1)}} \def\Hii{H^{(2)}} \def\Hk{\hat H^{(k)}}
\def\hi{H^{(i)}}
\def\Ll#1#2#3{\L_{#1}^{#2}\bigl(#3\bigr)}
\let\l\lambda \let\o\omega \def\bo{\overline\o}
\def\cho{\breve\o}
\def\wo{\widetilde\o}
\let\d\partial
\def\L{\bigwedge\nolimits}
\let\ints\cap \let\union\cup
\let\ph\varphi \let\ga\gamma \let\r\rho
\let\el\ell
\def\cq{,\quad}
\def\g{\frak g}
\def\n{{\cal N}}
\def\norm#1{\|#1\|}
%
%
\catcode`\@=11
\def\Eqalign#1{\null\,\vcenter{\openup\jot\m@th
	\ialign{\strut\hfil$\displaystyle{##}$&
 	$\displaystyle{{}##}$\hfil&&
  \quad\hfil$\displaystyle{##}$&
  $\displaystyle{{}##}$\hfil\crcr
 #1\crcr}}\,}
\catcode`\@=12
\def\Proof{\medbreak\noindent{\csc Proof}:\enspace}
\newif\ifprep\prepfalse
\parskip = 4pt plus 2pt minus 2pt
\widetildes
\widehats
\barrs

\proofmodefalse
\refsin

\font\csc=cmcsc10
\def\shorty#1#2{} \def\shortyear{\expandafter\shorty\the\year}
\ifprep
\rightline{\csc UCM--FTII Preprint
\the\month/\expandafter\shortyear}%
\fi
\Title
The Multidimensional Darboux Transformation.

\author
Artemio Gonz\'alez-L\'opez\\
Departamento de F\'\i sica Te\'orica II\\
Universidad Complutense\\
Madrid\\
SPAIN \quad 28040
\email artemio@ciruelo.fis.ucm.es
\support DGICYT Grants PB92--0197 and PB96--0197.

\author
Niky Kamran\\
The Fields Institute\\
222 College Street\\
Toronto, Ontario\\
CANADA \quad M5T 3J1
\email nkamran@math.mcgill.ca
\support an NSERC Grant{.} On
sabbatical leave from the Department of Mathematics and
Statistics, McGill University, Montreal, QC, H3A 2K6.

\printauthor
\dated
\Abstract A generalization of the classical one-dimensional
Darboux transformation to arbitrary $n$-dimensional oriented
Riemannian manifolds is constructed using an
intrinsic formulation based on the properties of twisted Hodge
Laplacians. The classical two-dimensional Moutard transformation is
also generalized to non-compact oriented Riemannian manifolds of
dimension $n \geq 2$. New examples of quasi-exactly solvable
multidimensional matrix Schr\"odinger operators on curved manifolds
are obtained by applying the above results.

\page
\Section i Introduction.
Our purpose in this paper is to define and study the properties
of a broad generalization to $n$ dimensions of the classical
Darboux transformation for Sturm-Liouville operators on the
line. Our approach will stem from a geometric generalization of
the basic intertwining relations underlying the classical
Darboux transformation to the context of certain twisted
Laplacians acting on the exterior algebra of an oriented
Riemannian manifold.

Let us begin by recalling the essentials of the one-dimensional
Darboux transformation,
\rf{Darboux}, \rf{Crum}. Consider a Sturm--Liouville operator
$h$, given by
$$
h=-\frac{d^2}{dx^2}+V(x),
$$
and let $e^{-\chi}$ be a nowhere vanishing eigenfunction of $h$
with eigenvalue $E_0$,
$$
(h-E_0)\,e^{-\chi} = 0.
$$
The classical Darboux transformation associates to $h$ the
Sturm--Liouville operator
$\htilde$ defined by
$$
\htilde=-\frac{d^2}{dx^2}+\Vtilde(x),
$$
where
$$
\Vtilde = V + 2\chi''.
$$
It is straightforward to verify that the operators $h$ and
$\htilde$, shifted by
$E_0$, can be factorized in the following way,
$$
h-E_0 = Q^+\,Q^-,\qquad \htilde-E_0 = Q^-\,Q^+,
$$
where $Q^+$ and $Q^-$ are first-order differential operators
defined by
$$
Q^+ = -\frac d{dx}+\chi',\qquad Q^- = \frac d{dx}+\chi'.
$$
The operators $h-E_0$, $\htilde-E_0$, $Q^+$ and $Q^-$ are
therefore related by the intertwining relations
$$
(h-E_0)\,Q^+ = Q^+\,(\htilde - E_0),\qquad Q^-\,(h - E_0) =
(\htilde-E_0)\,Q^-.
$$
We thus obtain a simple relation between the eigenfunctions
(formal or ${\rm L}^2$) of $h$ and those of its Darboux transform
$\htilde$. Indeed, if
$\psi$ is a formal eigenfunction of
$h$ with eigenvalue $E \neq E_0$, then it follows immediately
from the above intertwining relations that
$Q^-\psi$ will be a formal eigenfunction of $\htilde $ with the
same eigenvalue. Conversely, if
$\tilde \psi$ is a formal eigenfunction of
$\htilde$ with eigenvalue $E \neq E_0$, then
$Q^+\psi$ will be a formal eigenfunction of $h$ with eigenvalue
$E$. It is not difficult to show that this correspondence also holds
at the level of the
${\rm L}^2$ eigenfunctions of
$h$ and $\htilde$, so that the Darboux transformation
establishes a correspondence between bound states of $h$ and
$\htilde$.

It is well-known that the Darboux transformation plays an
important role in the theory of soliton solutions of integrable
evolution equations and in the method of inverse scattering,
\rf{DT}. It also provides a powerful
method for generating new exactly or quasi-exactly solvable
one-dimensional potentials from known ones \rf{Ush}. 
The Darboux transformation also appears as a basic tool in
the theory of special functions through the factorization
method of Infeld and Hull, \rf{IH}. The problem of extending the
Darboux transformation to the case of multi-dimensional
differential operators is therefore of considerable interest.

There are at least two natural candidates for what could be
called a Darboux transformation in two dimensions, namely the
{\it Laplace,}\/ \rf{Goursat}, and {\it Moutard,}\/
\rf{Moutard}, transformations. The Moutard transformation is
perhaps closer in spirit to the classical Darboux
transformation, since it is based on intertwining relations
analogous to the ones given above for the Darboux
transformation. This will be made explicit in Section 5. On the
other hand, the Laplace transformation plays a significant role
in the realm of integrable systems. For example, it naturally
gives rise to the Lax representation for the
$A_n$ Toda lattice, \rf{V}. It is also at the basis of some
important recent work on exactly solvable periodic
two-dimensional Schr\"odinger operators,
\rf{VN}. In any case, both the Laplace and Moutard
transformations preserve the class of linear elliptic
second-order differential operators in the plane. However,
unlike the one-dimensional Darboux transformation, they both
suffer from the major limitation that they yield only {\it 
one} \/eigenfunction for the transformed operator, \rf{VN}, the
reason being that {\it they do not incorporate the spectral
parameter $E$.}\/ Let us briefly illustrate this difficulty in
the case of the Laplace transformation. The conclusion is
analogous for the Moutard transformation.  Consider the
two-dimensional Schr\"odinger equation given by
$$
\left(-\frac{\d^2}{\d z \d \bar z}+V(x,y)\right)\psi =
E\, \psi, 
$$
in terms of complex coordinates $z=\frac12(x+iy)$, $\bar
z=\frac12(x-iy)$. Under the Laplace transformation, the wave
function $\psi$ gets mapped to the wave function $\hat\psi$
defined by
$$
\hat \psi = \frac{\d\psi}{\d \bar z},
$$
and it is easily verified that the transformed wave function
$\hat \psi$ satisfies the following Schr\"o\-ding\-er equation
for a particle in a magnetic field
$$
\left[-\frac{\d^2}{\d z \d \bar z}
+\frac{\d\log(V-E)}{\d\bar z}\,\frac\d{\d z}+V-E\right]
\hat\psi = 0.
$$
The explicit dependence of the coefficient of $\d/\d z$ on $E$
illustrates our point. There is also a drawback which is
specific to the Laplace transformation, namely that its
geometric generalization to $n$ dimensions, \rf{KT}, applies to
a class of highly overdetermined systems which bear no relation
to any natural spectral problem, although they are of course
interesting in their own right. Another essential limitation of
the Laplace and Moutard transformations is that they are only
defined for flat Laplacians expressed in Cartesian coordinates,
whereas many of the Schr\"odinger operators arising by symmetry
reduction involve curved Laplacians in very general coordinate
systems. One would therefore like to have a multi-dimensional
Darboux transformation which is defined in a {\it covariant}\/
way, which allows for {\it curvature}\/ of the underlying
manifold and which includes the spectral parameter in a natural
way. 

An indication on how to proceed is suggested by the work of
Andrianov, Borisov and Ioffe,
\rf{ABI}. In their scheme, one starts from a Schr\"odinger
operator
$h$ in the Euclidean plane, expressed in Cartesian coordinates,
and one constructs, starting from a nowhere vanishing eigenfunction
of $h$, the Moutard transform
$\htilde$ of
$h$, and a two-by-two matrix Schr\"odinger operator
$H$ which splits into the sum of two operators $\Hi$ and
$\Hii$. The intertwining relations between $h$, $\htilde$,
$\Hi$ and $\Hii$ imply that $h$ and $\Hi$ have the same
eigenvalues except possibly for the zero eigenvalue, and
similarly for
$\htilde$ and $\Hii$. (This construction was later generalized
in \rf{ABEI} to {\it flat}\/
$n$-dimensional Euclidean space, still using Cartesian
coordinates in an essential way.) After a careful analysis of this
scheme, we conclude that the operators $h$, $H$ and
$\htilde$ introduced in \rf{ABEI} can in fact be re-expressed as
twisted flat Laplacians acting on
$0$-forms, $1$-forms and $2$-forms, the latter being identified
with $0$-forms by means of the Hodge operator for the
underlying two-dimensional flat Euclidean metric. The
intertwining relations between
$h$,
$\htilde$, $\Hi$ and $\Hii$ will now follow immediately
from elementary properties of the twisted differentials and
codifferentials. 

Starting from this observation, we succeed in our
paper to construct a fully covariant and coordinate-free
multidimensional generalization of the classical Darboux
transformation, valid on an arbitrary curved
$n$-dimensional oriented Riemannian manifold.\/ Our
$n$-dim\-ensional Darboux transformation relates, via
intertwining relations involving twisted differentials and
codifferentials, the spectra and eigenfunctions of a string of
$n+1$ twisted Laplacians acting on
$k$-forms for $0
\leq k
\leq n$. It is noteworthy that these are the twisted Laplacians
which were used by Witten in his proof of the Morse
inequalities based on ideas from supersymmetry, \rf{WittenSMT}.
When expressed in any coordinate system, these twisted
Laplacians take the form of matrix Schr\"odinger operators
acting on the ${n \choose k}$ components of a $k$-form. In the
special case where the underlying manifold is flat Euclidean
space in Cartesian coordinates, our multidimensional Darboux
transformation reproduces the classical Darboux transformation
and the scheme of \rf{ABI} and \rf{ABEI}.

Our paper is organized as follows. In Section 2, we define the
twisted versions of the differential, the codifferential and
the Laplacian on forms. When expressed in local coordinates,
the latter will correspond to scalar and  matrix Schr\"odinger
operators on curved Riemannian manifolds. In Section 3, we
first derive the basic intertwining relations defining our
multidimensional Darboux transformation and give their spectral
interpretation. This generalization of the Darboux
transformation is valid for twisted Laplacians on an arbitrary
$n$-dimensional oriented Riemannian manifold. We then derive
the local coordinate expressions of the resulting matrix
Schr\"odinger operators in terms of the seed eigenfunction for
the original scalar Hamiltonian and the Riemann curvature of
the background metric. The connection between the spectra of
our matrix Hamiltonians admits an interesting interpretation in
terms of supersymmetry, cf.~\rf{WittenDBS}, \rf{WittenSMT},\rf{ABEI},
which we briefly recall. In Section 4, we show that our
multidimensional Darboux transformation reduces to the classical
Darboux transformation on the line and to a covariant coordinate-free
generalization to curved oriented Riemannian manifolds of the
scheme of references
\rf{ABI} and \rf{ABEI} in two dimensions. In Section
5, we derive an
$n$-dimensional generalization of the Moutard transformation
which applies to all twisted Laplacians. The spectral
interpretation of the multi-dimensional Moutard transformation
is significantly more limited than that of the multidimensional
Darboux transformation, because it only applies to the zero
modes of the twisted Laplacians. Finally, in Section 6 we
obtain new examples of multi-dimensional quasi-exactly or
exactly solvable matrix Schr\"odinger operators of physical
interest, of which very few seem to be known, by applying the
multidimensional Darboux transformation to various
quasi-exactly solvable planar Hamiltonians, \rf{GKO}.
%
\Section{not} Twisted Laplacians and Schr\"odinger operators.
Our purpose in this section is to define a twisted version of
the Laplacian on $k$-forms on an oriented Riemannian manifold.
On scalar functions, this twisted Laplacian will correspond to
a Schr\"odinger operator.

We start by setting up some notation. Let $M$ be an
$n$-dimensional oriented  Riemannian manifold, with metric
$(g_{ij})_{1\le i,j\le n}$. We shall denote by $\L^k(M)$ the 
vector space of differential
$k$-forms on $M$, and by
$\L(M) = \bigoplus_{k=o}^{n} \L^k(M)$ the exterior algebra of
differential forms on
$M$. In a local coordinate system 
$x=(x^1,\dots,x^n)$ a $k$-form 
$\o\in\L^k(M)$ will be written as
$$
\o = \frac1{k!}\,\o_{i_1\dots 
i_k}(x)\,dx^{i_1}\wedge\dots\wedge dx^{i_k},
\Eq{kform}
$$
where $\o_{i_1\dots i_k}$ is an antisymmetric tensor. Here, and
in what follows, we are using the summation convention.

Since $M$ is Riemannian and oriented, it admits a volume form \/
$\mu\in\L^n(M)$, which in a positively oriented coordinate 
system $x=(x^1,\dots,x^n)$ can be written as
$$
\mu = \sqrt{g(x)}\, dx^1\wedge\dots\wedge dx^n,
$$
where
$$
g = \det \bigl(g_{ij}\bigr).
$$
The Riemannian metric on 
$M$ induces a scalar product on the exterior algebra of $M$ in
a natural way.  Namely, if $\alpha$ and $\beta$ are two
$k$-forms with (antisymmetric) components $\alpha_{i_1\dots 
i_k}$ and $\beta_{i_1\dots i_k}$, respectively, one defines
$$
\inner\alpha\beta = \frac1{k!}\,\alpha_{i_1\dots 
i_k}\,\beta^{i_1\dots i_k}
= \frac1{k!}\,\alpha^{i_1\dots i_k}\,\beta_{i_1\dots i_k},
$$
where the functions
$$
\alpha^{i_1\dots i_k} = g^{i_1j_1}\cdots\
g^{i_kj_k}\,\alpha_{j_1\dots  j_k}
$$
are the contravariant components of $\alpha$, and
$$
(g^{ij}) = (g_{ij})^{-1}
$$
is the contravariant metric tensor. It  is clear that the
definition of the function $\inner\alpha\beta$ does not depend
on the local  coordinates used; hence one defines the inner
product $(\alpha,\beta)$ of the $k$-forms 
$\alpha,\beta\in\L^k(M)$ by
$$
(\alpha,\beta) = \int_M \inner\alpha\beta\,\mu.
$$
The inner product $(\alpha,\beta)$ will only be  defined if the
above integral is convergent; for instance, this will be  the
case if $\alpha$ and $\beta$ are compactly supported, or in
particular if $M$ is a compact manifold.

Given a $k$-form $\beta\in\L^k(M)$, the Hodge star
$*\beta\in\L^{n-k}(M)$ of
$\beta$ is defined as usual as  the unique $(n-k)$-form
satisfying
$$
\alpha\wedge*\beta = \inner\alpha\beta\,\mu,
\qquad\forall\alpha\in\L^k(M).
$$
The Hodge star operator
$*:\L^k(M)\to\L^{n-k}(M)$ has the following 
elementary properties, which we list here for
future reference:
$$
\openup4\jot\eeqn{
\alpha\wedge*\beta = 
\beta\wedge*\alpha,\qquad\alpha,\beta\in\L^k(M)\Eq{hodge1}\\
**\alpha =
(-1)^{k(n-k)}\,\alpha,\qquad\alpha\in\L^k(M)\Eq{hodge2}\\ 
*\mu = 1,\qquad *1 = \mu.
\Eq{hodge3}
}
$$
In local coordinates the 
components of the Hodge dual of a $k$-form $\alpha$ are
given by
$$
(*\alpha)_{j_{k+1}\dots j_n} = \frac{\sqrt
g}{k!}\,\epsilon_{j_1\dots  j_n}\alpha^{j^1\dots j^k},
\qquad\alpha\in\L^k(M),
\Eq{star}
$$
where $\epsilon_{j_1\dots j_n}$ is antisymmetric under
permutations  of its indices, and
$$
\epsilon_{12\dots n} = +1.
$$

The Hodge star operator is used to define the {\it 
codifferential}\/ $\delta$ of a $k$-form
$\alpha$ by the formula
$$
\delta\alpha = (-1)^{n(k-
1)+1}*d(*\alpha),\qquad\alpha\in\L^k(M).
$$
Thus, $\delta$ maps $\L^k(M)$ into $\L^{k-1}(M)$; in 
particular, $\delta=0$ on
$\L^0(M) = C^\infty(M)$. In local coordinates, one can show 
that
$$
(\delta\alpha)_{i_1\dots i_{k-1}} =
-\nabla^j\,\alpha_{ji_1\dots i_{k- 1}},
\Eq{cod}
$$
where $\nabla$ denotes the covariant derivative associated to
the  Riemannian metric. The odd looking sign in the definition
of $\delta$ is not arbitrary,  since with this choice of sign
the operators $d$ and $\delta$ are the formal adjoint of one 
another with respect to the scalar product on $\L(M)$, 
$$
d_k^\dagger = \delta_{k+1},\qquad 0\le k\le n-1.
$$
It also follows 
immediately from its definition that $\delta$ is
a coboundary operator, namely
$$
\delta^2 = 0.
$$
Furthermore, the local exactness of $d$ (Poincar\'e's lemma)
implies  that $\delta$ is also locally exact: in other words,
if $\delta\o = 0$ then there is an open neighborhood of every
point of $M$ on which 
$\o=\delta\alpha$, for some differential form
$\alpha$.

The {\it Laplacian}\/ $\Delta:\L(M)\to\L(M)$, defined by
$$
-\Delta = d\delta+\delta d,
$$
maps each space $\L^k(M)$ into itself for all $k=0,1,\dots,n$.
Moreover, from the elementary properties of the codifferential
it  follows that $-\Delta$ is (formally) self-adjoint and
non-negative:
$$
\openup2\jot\eqalign{
(\alpha,-\Delta\beta) &= (-
\Delta\alpha,\beta),\qquad\alpha,\beta\in\L^k(M)\cr
(\alpha,-\Delta\alpha) &= 
(d\alpha,d\alpha)+(\delta\alpha,\delta\alpha)\ge0,
\qquad\alpha\in\L^k(M).\cr
}
$$
If $f\in\L^0(M)$ is a smooth function on $M$, we obtain using
\eq{cod} that
$$
\Delta f = -\delta df = \nabla^i\nabla_i f = \frac1{\sqrt
g}\frac{\d}{\d  x^i}\left(\sqrt g g^{ij}\frac{\d f}{\d
x^j}\right)
\Eq{LB}
$$
is the classical Laplace--Beltrami operator on $\L^0(M)$.
If 
$\o$ is a $k$-form, the components of $-
\Delta\o$ in a local coordinate system are given by the 
following expression:
$$
\Eeqn{
(-\Delta\o)_{i_1\dots i_k} &= 
-\nabla^j\nabla_j\,\o_{i_1\dots i_k}+
\sum_{q=1}^k\,(-1)^{q+1}\,R^j{}_{i_q}\, \o_{ji_1\dots 
\widehat{i_q}\dots 
i_k}\cr
&\qquad{}+2\sum_{1\le p<q\le
k}\,(-1)^{p+q+1}\,R^j{}_{i_p}{}^l{}_{i_q}\,\o_{jl
i_1\dots\widehat{i_p}\dots\widehat{i_q}\dots i_k}.
\Eq{Lap}}
$$
In the above equation a hat over an index means that that index
is to be omitted,  the Riemann tensor $R^i{}_{jhl}$ is defined
by
$$
R^i{}_{jhl} = \frac{\d\Gamma^i_{jh}}{\d x^l}-
\frac{\d\Gamma^i_{jl}}{\d x^h}
+ \Gamma^i_{pl} \Gamma^p_{jh} - \Gamma^i_{ph} \Gamma^p_{jl}
$$
in terms of the connection coefficients $\Gamma^i_{jl}$ of the
metric, and $R^i{}_j$ is given in terms of the Ricci tensor
$R_{ij}$ by
$$
R^i{}_j = R^{ih}{}_{hj} = g^{ih}\,R_{hj}.
$$

A {\it Schr\"odinger operator}\/ (or {\it Hamiltonian}\/) on
$M$ is a second-order  differential operator
$h:\L^0(M)\to\L^0(M)$ of the form
$$
h = -\Delta+V,
\Eq{hdef}
$$
where $\Delta$ is the classical Laplace--Beltrami operator
\eq{LB},  and the {\it potential}\/ $V$ is a scalar function.
Just as the classical Laplace--Beltrami operator is  the
restriction of $-(d\delta+\delta d)$ to $\L^0(M)$, an arbitrary 
Schr\"odinger operator $h$ can be expressed in terms of a
``twisted" version of the latter operator.

To this end, given a smooth real-valued function $\chi$ on $M$
we introduce the {\it twisted differentials}
$$
d^\pm = e^{\pm\chi}\,d\,e^{\mp\chi}
$$
and the {\it twisted codifferentials}
$$
\delta^\pm = e^{\pm\chi}\,\delta\,e^{\mp\chi}.
$$
The operators $d^\pm$ and 
$\delta^\pm$ have the following properties, which follow
directly  from analogous properties of $d$ and $\delta$:

\medskip{\advance \leftskip by 4 em
\item{i)}\quad
$(d^\pm)^\dagger = \delta^\mp$
\item{ii)}\quad
$(d^\pm)^2 = (\delta^\pm)^2 = 0$
\item{iii)}\quad
$d^\pm$ and $\delta^\pm$ are locally exact\par}

\medskip\noindent There are two natural ways of ``twisting" the
operator $\Delta$ in  such a way that the resulting operator
maps $\L^k(M)$ into  itself for $k=0,1,\dots,n$, and is
formally self-adjoint and non-negative;  namely, we can define
the {\it twisted Laplacians}%
\footnote{${}^{\dagger}$}{The one-parameter family of twisted
Laplacians $H_t$ obtained from $H$ by setting $\chi=t\,\chi_0$,
where
$t\in\R$ and $\chi_0$ is a Morse function on $M$, was studied
by Witten in his derivation of the Morse inequalities using
supersymmetric quantum mechanics, cf.~\rf{WittenSMT}.}
$$
H = d^-\delta^+ + \delta^+d^-
$$
and
$$
\Ht = d^+\delta^- + \delta^-d^+.
$$

Let us take the first of these operators,  and apply it to a
function 
$f\in\L^0(M)$. We shall need the following lemma, whose  proof
is straightforward, to simplify the computation:
\Lemma{dfdg} If $f$ and $g$ are scalar functions on $M$,  we
have the  identity
$$
\delta(f\,dg) = - f\Delta g - \nabla f\cdot\nabla g.
$$

Here $\nabla f$ denotes the gradient of $f$, that is the vector
field  with components
$$
(\nabla f)^i = g^{ij}\nabla_j f = g^{ij}\frac{\d f}{\d x^j},
$$
and $\nabla f\cdot\nabla g = \inner{df}{dg}$. Using the lemma
we  obtain the following expression for the action of the
operator $H$ on  a function $f\in\L^0(M)$:
$$
\eqalign{
Hf &= \delta^+d^-f = e^\chi\delta\left[e^{-\chi}\cdot
e^{-\chi}d(e^\chi f)\right]\cr &=
e^\chi\delta(e^{-\chi}df)+e^\chi\delta(e^{-\chi}f d\chi)\cr &=
-\Delta f - e^\chi\nabla(e^{-\chi})\cdot\nabla f - f\Delta\chi-
e^\chi\nabla(e^{-\chi}f)\cdot\nabla\chi\cr &=
(-\Delta+V-E_0)\,f, }
$$
where
$$
V= (\nabla\chi)^2-\Delta\chi+E_0,
\Eq{V}
$$
and $E_0$ is an arbitrary real constant whose purpose will soon
become clear. (Note that $V$ is real, since $\chi$ is a
real-valued function.) Thus, an arbitrary Schr\"odinger operator
$h$ can be represented as
$$
h = H_0+E_0\equiv(\delta^+d^-+d^-\delta^+)_0+E_0,
$$
provided that the function $\chi$ satisfies \eq{V}.  The
meaning of \eq{V} and of the arbitrary constant $E_0$ is
apparent if we observe  that
$$
(h-E_0)e^{-\chi} = H_0e^{-\chi}=\delta^+d^-e^{-\chi} = \delta^+
\bigl[e^{-\chi} d(1)\bigr] = 0.
$$
In other words, \eq{V} is equivalent to the fact that
$e^{-\chi}$ is a  formal eigenfunction of $h$ with eigenvalue
$E_0$. Note that here,  and in what follows, by a {\it formal
eigenfunction} of an operator $h$ we  simply mean a solution
$\psi$ of the eigenvalue equation $(h-E)\psi=0$,  regardless of
the boundary conditions (like square integrability)  that may
be used to define true eigenfunctions of $h$.

Similarly, if we apply the twisted Laplacian $\Ht$ defined
above to a scalar function $f\in\L^0(M)$ we obtain
$$
\Ht f = \delta^- d^+ f = (-\Delta+\Vtilde-E_0)\,f
$$
with $\Vtilde$ given by
$$
\Vtilde = (\nabla\chi)^2+\Delta\chi+E_0 = V + 2\Delta\chi,
\Eq{Vt}
$$
since $\Ht$ is obtained from $H$ by replacing $\chi$ with
$-\chi$. Letting
$$
\htilde = -\Delta+\Vtilde
\Eq{htdef}
$$
we then have
$$
(\htilde-E_0) e^\chi = 0.
$$
Note that the Schr\"odinger operator $\htilde$ is the {\it
Moutard transform}\/ of $h$,
\rf{Moutard}.
\Section{mdt} The Multidimensional Darboux Transformation.
The goal of this section is to define the multidimensional
Darboux transformation. The definition will be naturally
suggested by some fundamental intertwining relations involving
the differential operators on $\L(M)$ introduced in the
previous section.

Let us consider the operator $H:\L(M)\to\L(M)$, which we shall
decompose as follows:
$$
H = \Hi + \Hii,
$$
where
$$
\Hi = d^- \delta^+,\qquad \Hii = \delta^+ d^-.
$$
Observe that, since $\delta^+$ and $d^-$ are coboundary
operators, we have
$$
\Hi\,\Hii = \Hii\,\Hi = 0.
\Eq{h120}
$$
Since the operator $H$ maps $\L^k(M)$ into itself for
$k=0,1,\dots,n$, we can also write
$$
H = \bigoplus_{k=0}^n H_k,
$$
where $H_k$ is the restriction of $H$ to $\L^k(M)$. The operator
$H_k$ can be decomposed as
$$
H_k = \Hi_k + \Hii_k,
$$
where $\Hi_k$ and $\Hii_k$ are the restrictions of $\Hi$ and
$\Hii$ to
$\L^k(M)$, namely
$$
\Hi_k = d^-_{k-1} \delta^+_k,\qquad
\Hii_k = \delta^+_{k+1} d^-_k.
$$
Notice that
$$
\Hi_0 = \Hii_n = 0.
$$
From \eq{h120} we obtain
$$
\Hi_k\, \Hii_k = \Hii_k\, \Hi_k = 0.
\Eq{hk120}
$$
By construction, $H$ is formally self-adjoint and non-negative.
Therefore the same is true for the operators
$H_k:\L^k(M)\to\L^k(M)$ for
$k=0,1,\dots,n$. Likewise, both $\Hi_k$ and $\Hii_k$ are
formally self-adjoint and non-negative, the latter property
being a consequence of the identities
$$
\eqalign{
(\alpha,\Hi_k\alpha) &= (d^-\alpha,d^-\alpha),\cr
(\alpha,\Hii_k\alpha) &= (\delta^+\alpha,\delta^+\alpha),
\qquad\alpha\in\L^k(M),\quad k=0,1,\dots,n.
}
$$

The following {\it intertwining relations} are an immediate
consequence of the definition of $\Hi$ and $\Hii$:
$$
\openup2\jot
\eeqn{
\delta^+\,\Hi = \Hii\,\delta^+ \\
\Hi\,d^- = d^- \Hii.\\
\noalign{\noindent\rm These imply}
\delta^+_{k+1}\,\Hi_{k+1} = \Hii_k\,\delta^+_{k+1} 
\Eq{int1}\\
\Hi_{k+1}\,d^-_k = d^-_k \Hii_k,\rlap{\qquad
$k=0,1,\dots,n-1$.}\Eq{int2}
}
$$
The intertwining relations \eq{int1}--\eq{int2} have important
consequences for the spectra of the operators $\Hi_{k+1}$ and
$\Hii_k$, that we shall now explore.

\Pr{int}
\item{i\/\rm)} If $\o\in\L^{k+1}(M)$ is an eigenform of
$\Hi_{k+1}$ with eigenvalue
$\l\ne0$, then
$\delta^+\o\equiv\delta^+_{k+1}\o$ is an eigenform of
$\Hii_k$ with the same eigenvalue.
\item{ii\/\rm)} Likewise, if $\o\in\L^k(M)$ is an eigenform of
$\Hii_k$ with eigenvalue
$\l\ne0$, then
$d^-\o\equiv d^-_k\o$ is an eigenform of $\Hi_{k+1}$ with the
same eigenvalue.\endgraf\par
\Proof The only point that is not an immediate consequence of
the intertwining relations is that $\delta^+\o$ in part i)
cannot vanish identically, and likewise for
$d^-\o$ in part ii). Let us show, for instance, that
$\delta^+\o\ne0$ in part i). If $\delta^+\o=0$, then by the
local exactness of $\delta^+$ (property iii) in \section{not})
for every $p\in M$ there is an open neighborhood $U_p$ of $p$
such that $\o = \delta^+\alpha_p$ on $U_p$, for some
differential form $\alpha_p$ on $U_p$.  But this would imply
that
$\Hi_{k+1}\o = d^-\delta^+(\delta^+\alpha_p) = 0$ on $U_p$, for
every $p\in M$, so that $\Hi_{k+1}\o=0$ on $M$. Hence $\l=0$,
contradicting the hypothesis.\qed

\bigskip
Let $L$ be a differential operator on $\L(M)$ mapping each
subspace $\L^k(M)$ into itself. Examples of such an operator are
$H$, $\Hi$ and $\Hii$. We shall denote by $\CA(L)\subset\L(M)$
the set of {\it admissible forms}\/ for $L$, that is the
subspace of the space of square-integrable forms
$L^2\bigl(\L(M)\bigr)$ satisfying any additional boundary or
asymptotic conditions that are appropriate for the problem
being considered. The set of admissible forms for $L_k$ is then
$\CA(L_k) \equiv \CA_k(L) = \CA(L)\ints\L^k(M)$. The {\it spectrum}\/ of
$L$,
denoted by $\sigma(L)$, is the set of numbers $\l$ such
that there is an admissible non-zero eigenform
$\o\in\CA(L)$ satisfying the eigenvalue equation
$L\,\o=\l\,\o$.  We shall also use the
convenient notation
$$
\sigma'(L) = \sigma(L)-\{0\}.$$
The linear space of all $k$-forms $\o$ satisfying
the eigenvalue equation $L\,\o=\l\,\o$ will be denoted by $\Ll\l kL$.
Equivalently,
$\Ll\l kL$ is the set of eigenforms of $L_k$ with eigenvalue
$\l$, together with the zero $k$-form.

By the formal self-adjointness and non-negativity of
$H_k$, $\Hi_k$ and $\Hii_k$, the eigenvalues of these operators
are real and non-negative. \pr{int} has the following immediate
corollary:

\Co{spec} If the operators $\delta^+$ and $d^-$ map
$\CA\equiv\CA(H)$ into itself, then the spectra of $\Hii_k$ and
$\Hi_{k+1}$ are related by
$$
\sigma'(\Hii_k) = \sigma'(\Hi_{k+1}),\qquad
k=0,1,\dots,n-1.
\Eq{hkiii}
$$

In the remainder of this section, we shall not distinguish
between true and formal eigenforms, unless otherwise indicated.
The following lemma, whose proof is elementary, will have
non-trivial consequences in what follows:

\Lm{LA} Let $V$ be a vector space, and let $L:V\to V$ be a
linear operator. Suppose that
$L=L_1+L_2$, with $L_1\,L_2 = L_2\,L_1 = 0$. The following
statements are then true:
\item{i\/\rm)} If $v$ is an eigenvector of $L$ with eigenvalue
$\l$, either
$L_i v=\l\,v$ and $L_j v = 0$ for some $i,j\in\{1,2\}$ with
$i\ne j$, or
$L_i v$ is an eigenvector of
$L_i$ with eigenvalue $\l$ for $i=1,2$.
\item{ii\/\rm)} $\sigma(L) \subset
\sigma(L_1)\union\sigma(L_2).$
\item{iii\/\rm)}
$\sigma'(L) = \sigma'(L_1)\union\sigma'(L_2).$

From the previous lemma and \eq{hk120}, it follows that
$$
\sigma'(H_k) = \sigma'(\Hi_k)\union\sigma'(\Hii_k).
\Eq{sphk}
$$
Thus, the spectrum of $H_k$, with the possible exception of the
zero eigenvalue, is simply the union of the spectra of its
components $\Hi_k$ and
$\Hii_k$. The spectra of $\Hi_{k+1}$ and $\Hii_k$
($k=0,1,\dots,n-1$) are identical, except perhaps for the zero
eigenvalue. The operator $\delta^+$ maps eigenforms of
$\Hi_{k+1}$ with non-zero eigenvalue into eigenforms of
$\Hii_k$ with the same eigenvalue, and similarly $d^-$ maps
eigenforms of $\Hii_k$ with non-zero eigenvalue into eigenforms
of $\Hi_{k+1}$ with the same eigenvalue. From the identity
\eq{hkiii} we easily obtain the following relation between the
even and odd components of $H$:
$$
\sigma'\biggl(\bigoplus_{j=0}^{[n/2]}H_{2j}\biggr) =
\sigma'\biggl(\bigoplus_{j=0}^{\left[\frac{n-1}2\right]}H_{2j+1}\biggr).
$$

\Df{darboux} Let $\l\ne0$, and $k=0,1,\dots,n-1$.
If $\o\in\Ll\l{k+1}{\Hi}$, its {\it Darboux transform}
is the $k$-form $\delta^+\o\in\Ll\l{k}{\Hii}$. Similarly, the
Darboux transform of a $k$-form $\o\in\Ll\l{k}{\Hii}$
is the $(k+1)$-form $d^-\o\in\Ll\l{k+1}{\Hi}$.

From the definition of $\hi$ it directly follows that if
$\l\ne0$ the restriction
$\delta^+:\Ll\l{k+1}{\Hi}\to\Ll\l{k}{\Hii}$ is invertible, its
inverse being the restriction
$\l^{-1}\,d^-:\Ll\l{k}{\Hii}\to\Ll\l{k+1}{\Hi}$.

The Darboux transformation we have just defined acts in a
natural way on eigenforms of the partial Hamiltonians $\Hi$ and
$\Hii$. In the same spirit, we shall see next how to use the
Darboux transformation to construct new eigenforms of $H$ of
degree $k-1$ and/or $k+1$ starting from a given eigenform of
degree
$k$ with non-zero eigenvalue. To this end, let $\o\in\L^k(M)$
be an eigenform of
$H$ with eigenvalue
$\l\ne0$. By \lm{LA}, either $\hi\,\o = \l\,\o$ for some
$i\in\{1,2\}$, or
$\hi\,\o$ is an eigenform of $\hi$ with eigenvalue $\l\ne0$ for
$i=1,2$. In the first case, the Darboux transform of $\o$ is
well defined according to
\df{darboux}, and belongs to either $\Ll\l{k-1}{\Hii}$ (when
$i=1$), or to
$\Ll\l{k+1}{\Hi}$ (when $i=2$). By \lm{LA}, it follows that the
Darboux transform of $\o$ is an eigenform of $H$ in this case.
In the second case, the Darboux transforms of both $\Hi\o$ and
$\Hii\o$ are defined and, as before, are eigenforms of $H$ with
eigenvalue $\l$. Hence in this case both
$\delta^+\Hi\,\o\equiv\delta^+d^-\delta^+\,\o$ and
$d^-\Hii\,\o\equiv d^-\delta^+d^-\,\o$ are eigenforms of $H$
with eigenvalue
$\l$ and degree equal to
$k-1$ and $k+1$, respectively. Note that in this case both
$\Hi\o$ and
$\Hii\o$ are also eigenforms of $H$ of degree $k$ with
eigenvalue $\l$, neither of which is proportional to $\o$
(although the span of $\o$, $\Hi\o$ and
$\Hii\o$ is obviously two-dimensional). Thus in the second
case, i.e, when
$\hi\,\o
\ne 0$ for
$i=1,2$, we can construct {\it three}\/ new eigenforms of
$H$ of degrees $k-1$, $k$ and $k+1$ and eigenvalue $\l\ne0$
starting from a known eigenform $\o\in\Ll\l kH$.

The above construction could have been carried out using the
twisted Laplacian
$\Ht$ instead of $H$. However, we shall now show that the two
constructions are equivalent:

\Pr{eqhht} For $k=0,1,\dots,n$, the operators $H_{n-k}$ and
$\Ht_k$ are linearly equivalent under Hodge duality:
$$
\Ht_k = (*)^{-1} H_{n-k} *.
\Eq{hth}
$$

\Proof
An elementary calculation shows that
$$
* H_{n-k} * = (-1)^{k(n-k)} \Ht_k,
$$
from which \eq{hth} follows using \eq{hodge2}.\qed

In other words, if we identify the space $\L^{n-k}(M)$ to
$\L^k(M)$ under the Hodge star operator, $H_{n-k}$ is
transformed into $\Ht_k$. In particular,
$H_n$ is equivalent under this identification to
$\Ht_0=\htilde-E_0$, the Moutard transform of $H_0=h-E_0$.

We end this section by deriving local coordinate expressions
for the multidimensional Darboux transformation and for the
component Hamiltonians
$H_k$ and
$\Htilde_k$. By definition,
$$ d^-\o = e^{-\chi} d(e^\chi\o) = d\o + d\chi\wedge\o,
$$ so that if $\o\in\L^k(M)$ we have the following local
coordinate expression
$$
\eeqn{
(d^-\o)_{ji_1\dots i_k} = (\nabla_{[j}+\chi_j)\,\o_{i_1\dots
i_k]}\\ = (\nabla_j+\chi_j)\,\o_{i_1\dots
i_k}-\sum_{r=1}^k(\nabla_{i_r}+\chi_{i_r})\,\o_{i_1\dots
i_{r-1}ji_{r+1}\dots i_k}.
\Eq{dm} }
$$
Here the square brackets denote antisymmetrization, and
$\chi_j=\nabla_j\chi=\d\chi/\d x^j$. Similarly, using \eq{cod}
we obtain
$$
(\delta^+\o)_{i_1\dots i_{k-1}} =
(-\nabla^j+\chi^j)\,\o_{ji_1\dots i_{k-1}},
\qquad\o\in\L^k(M).
\Eq{deltap}
$$
Using the local formulas \eq{dm} and \eq{deltap} we obtain,
after a straightforward calculation
$$
\Eeqn{
(H_k\o)_{i_1\dots i_k} &= 
\left[-\nabla_j
\nabla^j+(\nabla\chi)^2-\Delta\chi\right]\,\o_{i_1\dots i_k}\\
&{}+\sum_{r=1}^k
\left\{\nabla^j,\nabla_{i_r}\right\}\chi\cdot\o_{i_1\dots
i_{r-1}ji_{r+1}\dots i_k}\\
&{}+\sum_{r=1}^k\left[\nabla^j,\nabla_{i_r}\right] \o_{i_1\dots
i_{r-1}ji_{r+1}\dots i_k}.
\Eq{hki}}
$$
This formula can be further simplified as follows. In the first
place, note that since
$\chi$ is a scalar function we have
$$
\left\{\nabla^j,\nabla_{i_r}\right\}\chi =
2\,\nabla^j\nabla_{i_r}\chi = 2\,\nabla_{i_r}\nabla^j\chi.
$$
Secondly, from the identity
$$
\left[\nabla_j,\nabla_l\right]\o_{i_1\dots i_k} =
\sum_{r=1}^k\,R^h{}_{i_r j l}\,\o_{i_1\dots
i_{r-1}hi_{r+1}\dots i_k}
$$
we obtain, after straightforward manipulations
$$
\eqalign{
\sum_{r=1}^k\left[\nabla^j,\nabla_{i_r}\right] \o_{i_1\dots
i_{r-1}ji_{r+1}\dots i_k} &=
\sum_{r=1}^k\,(-1)^{r+1}\,
R^h{}_{i_r}\,\o_{hi_1\dots\widehat{i_r}\dots i_k}\cr &{}+
\sum_{1\le r<q\le
k}\,(-1)^{r+q+1}\,R^j{}_{i_r}{}^h{}_{i_q}\,
\o_{jhi_1\dots\widehat{i_r}\dots\widehat{i_q}\dots
i_k}.}
$$
Substituting the previous formulas into \eq{hki} we get the
following coordinate expression for the action of $H_k$:
$$
\Eeqn{
(H_k\o)_{i_1\dots i_k} &= 
\left[-\nabla_j
\nabla^j+(\nabla\chi)^2-\Delta\chi\right]\,\o_{i_1\dots i_k}\\
&{}+2\sum_{r=1}^k\,\nabla^j\nabla_{i_r}\chi\cdot\o_{i_1\dots
i_{r-1}ji_{r+1}\dots i_k}\\ &{}+\sum_{r=1}^k\,(-1)^{r+1}\,
R^h{}_{i_r}\,\o_{hi_1\dots\widehat{i_r}\dots i_k}\\
&{}+\sum_{1\le r<q\le
k}\,(-1)^{r+q+1}\,R^j{}_{i_r}{}^h{}_{i_q}\,
\o_{jhi_1\dots\widehat{i_r}\dots\widehat{i_q}\dots
i_k}.
\Eq{hk}
}
$$
When $\chi=0$ the previous expression reduces to the formula
for minus the Laplacian, eq.~\eq{Lap}. In particular, the
operator $H_k$ has the structure
$$
H_k = -\Delta_k + V_k,
$$
where $\Delta_k$ is the restriction of the Laplacian to
$\L^k(M)$, and $V_k$ acts like a matrix potential on the
components of any $k$-form:
$$
\eeqn{
(V_k\o)_{i_1\dots i_k} = 
\left[(\nabla\chi)^2-\Delta\chi\right]\,\o_{i_1\dots i_k}
+2\sum_{r=1}^k\,\nabla^j\nabla_{i_r}\chi\cdot\o_{i_1\dots
i_{r-1}ji_{r+1}\dots i_k}\\ = 
(V-E_0)\,\o_{i_1\dots i_k}
+2\,\sum_{r=1}^k\nabla^j\nabla_{i_r}\chi\cdot\o_{i_1\dots
i_{r-1}ji_{r+1}\dots i_k}.}
$$
The local coordinate expressions for the component Hamiltonians
$\Ht_k$ are obtained from the previous formulas by replacing
$\chi$ with $-\chi$.

As first pointed out by Witten, \rf{WittenDBS}, \rf{WittenSMT},
\rf{ABEI}, the connection between the spectra of the component
Hamiltonians $H_k$ discussed in this section admits an
interesting interpretation in terms of supersymmetry. Indeed,
the $n+1$ homogeneous components $\o_k$ of degree $k$ of a
differential form $\Omega = \bigoplus_{k=0}^n\o_k$ can be
interpreted as the components of a supermultiplet, with $k$-forms
regarded as being bosonic or fermionic depending on whether $k$ is
even or odd, respectively. The supercharges $Q^\pm$ are by
definition the operators
$$
Q^- = \delta^+,\qquad Q^+ = (Q^-)^\dagger = d^-,
$$
while the supersymmetric Hamiltonian
$$
H = \{Q^+,Q^-\}
$$
is just the twisted Laplacian. The remaining commutation
relations defining the standard supersymmetry algebra
$$
\{Q^\pm,Q^\pm\} = [Q^\pm,H] = 0
$$
hold thanks to the elementary properties of $d^\pm$,
$\delta^\pm$.

We can also introduce fermion creation and annihilation
operators
$$
b^-_i = \frac\d{\d x^i}\interior{},\qquad
b^{i+} = dx^i\wedge\,,
$$
where ${}\interior{}$ denotes the inner product. The usual
fermionic anticommutation relations $$
\{b^{i+},b^{j+}\} = \{b_i^-,b_j^-\} = 0,\qquad
\{b^{i+},b_j^-\} = \delta_j^i,
$$
as well as the identity
$$
b^{i+} = (b^{i-})^\dagger \equiv (g^{ij}\,b_j^-)^\dagger,
$$
follow easily from well known exterior algebra identities. The
supercharges $Q^\pm$ can be expressed in terms of the creation
and annihilation operators as
$$
Q^\pm = q^{i\mp}\,b_i^\pm,
$$
where
$$
q_i^\pm \o =
\frac1{k!}\,(\mp\nabla_i+\nabla_i\chi)\,\o_{i_1\dots i_k}\cdot
dx^{i_1}\wedge\dots\wedge dx^{i_k}
 =      g_{ij}\,q^{j\pm} \o.
$$
One can easily check the identity
$$
H = -\Delta + (\nabla\chi)^2 +
\nabla_i\nabla_j\chi\cdot[b^{i+},b^{j-}],
$$
which generalizes formula (13) of \rf{WittenSMT} (where an
orthonormal basis of the tangent space is used to define
creation and annihilation operators).
\Section{low} The Darboux Transformation in Low Dimensions.
Let us begin by showing that in the one-dimensional case the
multidimensional Darboux transformation reduces to the
classical Darboux transformation. Indeed, in this case $M=\R$,
$H_0=h-E_0$, and $H_1$ is equivalent to $\Ht_0=\htilde-E_0$
under the Hodge duality. If
$$h=-\frac{d^2}{dx^2}+V(x),$$ then from \eq{Vt} we have
$$\htilde=-\frac{d^2}{dx^2}+\Vtilde(x),$$
where
$$
\Vtilde = V + 2\chi''
$$
and
$$
(h-E_0)\,e^{-\chi} = 0.
\Eq{hchi}
$$
The operator $d^-$ maps eigenfunctions of $h=H_0+E_0$ with
eigenvalue
$E\ne E_0$ into eigenfunctions of $H_1$ with eigenvalue
$E-E_0$. Therefore, the operator
$Q^-=*d^-$ will map eigenfunctions of $h$ with eigenvalue $E\ne
E_0$ into eigenfunctions of
$\htilde$ with the same eigenvalue. If $x$ is a cartesian
coordinate we easily obtain
$$
Q^-\psi = *d^-\psi = (\frac d{dx}+\chi')\psi,
$$
so that
$$
Q^- = \frac d{dx}+\chi'.
\Eq{qm}
$$
Similarly, if $\phi$ is an eigenfunction of $\htilde=\Ht+E_0$
with eigenvalue $E$ then
$*\phi$ will be an eigenfunction of $H_1$ with eigenvalue
$E-E_0$, so that
$\delta^+(*\phi)$ is an eigenfunction of $h$ with eigenvalue
$E$. Proceeding as before we have
$$
\delta^+(*\phi) = (-\frac d{dx}+\chi')\phi,
$$
so that the operator
$$
Q^+ = -\frac d{dx}+\chi'
\Eq{qp}
$$
maps eigenfunctions of $\htilde$ with eigenvalue $E\ne E_0$
into eigenfunctions of
$h$ with the same eigenvalue. From the properties of $d^\pm$
and $\delta^\pm$, it follows that $Q^-$ and $Q^+$ are formally
the adjoint of one another under the Euclidean scalar product
on $\R$, and that
$$
h-E_0 = Q^+\,Q^-,\qquad \htilde-E_0 = Q^-\,Q^+.
\Eq{qpm}
$$
Equations \eq{hchi}, \eq{qm}, \eq{qp} and \eq{qpm} express the
classical Darboux transformation.

The two-dimensional Darboux transformation generalizes to
curved oriented surfaces the Darboux transformation introduced
in \rf{ABI} for $\R^2$ in cartesian coordinates. Indeed, let
$M$ be a two-dimensional oriented Riemannian manifold. The
component Hamiltonians are in this case $H_0=h-E_0$,
$H_1:\L^1(M)\to\L^1(M)$ and
$H_2:\L^2(M)\to\L^2(M)$, which is equivalent under the Hodge
duality to
$\Ht_0=\htilde-E_0$. The scalar Hamiltonians
$h$ and $\htilde$ are given by eqs.~\eq{hdef}, \eq{V},
\eq{htdef} and \eq{Vt}, while the function $e^{-\chi}$ as usual
satisfies $(h-E_0)\,e^{-\chi}=0$. To the operator $H_1$ acting
on one-forms there corresponds an operator $\hat H_1$ acting on
vector fields, defined by
$$
\hat H_1 X = (H_1 X^\flat)^\sharp,
\Eq{xflat}
$$
where $X^\flat = g_{ij}X^j dx^i$ is the one-form associated to
the vector field $X=X^i\d/\d x^i$, and $\sharp=\flat^{-1}$.
Using \eq{hk} we easily find the following local coordinate
expression for $\hat H_1X$:
$$
(\hat H_1 X)^i = \left[-\nabla_j
\nabla^j+(\nabla\chi)^2-\Delta\chi\right] X^i +
2\,\nabla^i\nabla_j\chi\cdot X^j + R^i_j\, X^j.
$$
From the well-known two-dimensional identity
$$
R^i_j = K\,\delta^i_j,
$$
where $K$ is the Gaussian curvature of $M$, we obtain the
equivalent expression
$$
(\hat H_1 X)^i = \left[-\nabla_j
\nabla^j+(\nabla\chi)^2-\Delta\chi+K\right]X^i +
2\,\nabla^i\nabla_j\chi\cdot X^j.
\Eq{h1}
$$
In flat space and cartesian coordinates, the above formula for
$\hat H_1$ reduces to formula (9) of \rf{ABI}. By \co{spec} and
eq.~\eq{sphk}, the spectra of the Hamiltonians
$h$,
$\htilde$ and $\hat H_1$ are related by
$$
\sigma'(\hat H_1) = \sigma'(h-E_0)\union\sigma'(\htilde-E_0).
$$

We shall now derive an expression for the two-dimensional
Darboux transformation in local coordinates. Let the operators
$q_i^\pm$ and $p_i^\pm$ be defined in local coordinates by
$$
\openup3\jot
\eeqn{
q_i^\pm = \mp\nabla_i+\chi_i = \mp e^{\pm\chi}\nabla_i e^{\mp\chi}\\
p_i^\pm = \sqrt g\,\epsilon_{ij} q^{j\mp}.
\Eq{qppm}}
$$ 
In particular, notice that the operators $q_i^\pm$
(resp.~$p_i^\pm$) transform like the components of a covariant
tensor (resp.~pseudo-tensor) of rank one under changes of local
coordinates.

If $\psi$ is an eigenfunction of $h$ with eigenvalue $E\ne
E_0$, then
$d^-\psi$ is an eigenfunction of $H_1$ with eigenvalue
$E-E_0$.  Using the general formula \eq{dm} we have
$$
(d^-\psi)_i = q_i^- \psi.
\Eq{dpsi}
$$
From \eq{xflat} it follows that the vector field with components
$$
g^{ij}q_j^-\psi \equiv q^{j-}\psi
\Eq{qup}
$$
is an eigenvector of $\hat H_1$ with eigenvalue $E-E_0$.
Suppose now that $\phi$ is an eigenfunction of $\htilde$ with
eigenvalue $E\ne E_0$. Then $*\phi$ is an eigenform of $H_2$
with eigenvalue $E-E_0$, so that
$\delta^+(*\phi)$ is an eigenform of $H_1$ with eigenvalue
$E-E_0$. From the local formulas \eq{star} and \eq{deltap} and
the fact that the pseudo-tensor with components
$\sqrt g\,\epsilon_{ij}$ is covariantly constant, it follows
that
$$
\delta^+(*\phi)_i = -p_i^-\phi.
$$
Therefore the vector field with components
$$
g^{ij}p_j^-\phi \equiv p^{i-}\phi
\Eq{pmphi}
$$
is an eigenvector of $\hat H_1$ with eigenvalue $E-E_0$. Note also
that $p^{i\pm}$ can be expressed as
$$
p^{i\pm} = \frac1{\sqrt g}\sum_j\epsilon_{ij}\,q_j^\mp.
$$

Conversely, if $\Psi=\psi_i\,dx^i$ is an eigenform of
$H_1^{(1)}$ with eigenvalue
$\l\ne0$, then (\lm{LA}) $\Psi$ is an eigenform of $H_1$ with
the same eigenvalue, and
$\delta^+\Psi$ is an eigenfunction of $h$ with eigenvalue
$\l+E_0$. From \eq{cod} we obtain
$$
\delta^+\Psi = -e^\chi\nabla^i(e^{-\chi}\psi_i) = q^{i+}\psi_i
= q^+_i\psi^i,
\Eq{qpsi}
$$
where the vector field $\Psi^\sharp=\psi^i\d/\d x^i$ is an
eigenvector of
$\hat H_1$. Likewise, if $\Psi$ is an eigenform of $H_1^{(2)}$
with eigenvalue
$\l\ne0$ then $\Psi$ is an eigenform of $H_1$ with the same
eigenvalue, and $\delta^+\Psi$ is an eigenfunction of $H_2$
with eigenvalue $\l+E_0$. It follows that $*\delta^+\Psi$ is an
eigenfunction of $\htilde$ with eigenvalue $\l+E_0$. Using the
local coordinate formulas \eq{star} and \eq{dm} we easily obtain
$$
*d^-\Psi = p_i^+\psi^i,
\Eq{ppsi}
$$
where again the functions $\psi^i$ are the components of an
eigenvector of $\hat H_1$ with eigenvalue $\l$. As before, in
flat space and cartesian coordinates eqs.~\eq{qppm},
\eq{qup}--\eq{ppsi} reduce to the corresponding formulas in
\rf{ABI}.

The above formulas expressing the two-dimensional Darboux
transformation in terms of the operators $q_i^\pm$ and
$p_i^\pm$ suggest that the component Hamiltonians and the
intertwining relations can also be written in terms of the
latter operators.  A straightforward computation using the
local coordinate expressions for $*$, $d$ and
$\delta$ shows that this is indeed the case. More precisely, we
have
$$
h-E_0 = q_i^+ q^{i-},\qquad
\htilde- E_0 = q_i^- q^{i+} = p_i^+ p^{i-}
$$
and $\hat H_1 = \hat H_1^{(1)} + \hat H_1^{(2)}$, with
$$
\bigl(\hat H_1^{(1)}\Psi\bigr)^i = q^{i-}q^+_j\psi^j\equiv
\bigl(\hat H^{(1)}\bigr)^i{}_j\psi^j,\qquad
\bigl(\hat H_1^{(2)}\Psi\bigr)^i = p^{i-}p^+_j\psi^j\equiv
\bigl(\hat H^{(2)}\bigr)^i{}_j\psi^j.
$$
Notice that, strictly speaking, $\bigl(\hat
H^{(k)}\bigr)^i{}_j$ is not the
$(i,j)$-th matrix element of the operator $\Hk$, since
$\nabla_i\psi^j$ in general depends on all the components of
$\Psi$. Similarly, the intertwining relations
\eq{int1}--\eq{int2} can be written as
$$
\openup2\jot
\eeqn{
q_i^+\bigl(\hat H^{(1)}\bigr)^i{}_j = (h-E_0)\,q^+_j\\
\bigl(\hat H^{(1)}\bigr)^i{}_jq^{j-} = q^{i-}(h-E_0)\\
p_i^+\bigl(\hat H^{(2)}\bigr)^i{}_j = (\htilde-E_0)\,p^+_j\\
\bigl(\hat H^{(2)}\bigr)^i{}_jp^{j-} = p^{i-}(\htilde-E_0).
}
$$
In flat space and cartesian coordinates, the above expressions
reduce to the corresponding ones in \rf{ABI}.

We shall now show how the classical Moutard transform,
\rf{Moutard}, is generalized in the case of an oriented
Riemannian surface. To this end, suppose that $\psi$ is a
formal eigenfunction of
$h$ with the same eigenvalue $E_0$ as $e^{-\chi}$. Note that
$\psi$ need not be proportional to $e^{-\chi}$, since we are
dealing with formal eigenfunctions. A {\it Moutard transform}\/
of $\psi$ is any function
$\pt$ satisfying
$$
d^+\pt = \delta^-(*\psi).
\Eq{mout}
$$
Locally, the above equation has a solution if and only if
$(h-E_0)\psi = 0$. Indeed, by the local exactness of $d^+$ the
compatibility condition of
\eq{mout} is
$$
d^+\delta^-(*\psi) = 0.
$$
The Hodge dual of the latter equation is simply
$$
\delta^+d^-\psi = (h-E_0)\psi = 0,
$$
as claimed. Note that the Moutard transform $\pt$ is locally
defined by
\eq{mout} only up to a constant multiple of $e^\chi$. Indeed,
if $\pt_1$ and
$\pt_2$ are two Moutard transforms of $\psi$ then their
difference satisfies
$d^+(\pt_1-\pt_2) = 0$, that is
$d\left[e^{-\chi}(\pt_1-\pt_2)\right] = 0$.

If $\pt$ is any Moutard transform of $\psi$, then
$$
(\htilde-E_0)\pt = 0,
$$
where as before $\htilde$ is the Moutard transform of $h$.
Indeed, applying $\delta^-$ to \eq{mout} we obtain
$$
0 = \delta^-d^+(\pt) = \Ht_0\pt = (\htilde-E_0)\pt.
$$
\Section{moutard} The Multidimensional Moutard Transformation.
We shall derive in this section a generalization to oriented
Riemannian manifolds of arbitrary dimension of the classical
two-dimensional Moutard transformation introduced in the
previous section. The key to this generalization is a
remarkable connection between the zero eigenspaces of the
operators $\Hii_k$ and $\Hi_{k+2}$ (or, equivalently, of the
components $\Htii_{k}$ and $\Hti_{k-2}$ of
$\Ht$ defined below) that we shall describe next.

Throughout this section, $M$ will denote an oriented Riemannian
manifold of dimension $n\ge2$ with trivial de Rham cohomology.
In particular, the latter condition will always hold if we
restrict ourselves to a contractible open subset of $M$.
Let us decompose $\Ht$ as
$$
\Ht = \Hti+\Htii,
$$
with
$$
\Hti = \delta^-d^+,\qquad \Htii = d^+\delta^-,
$$
so that
$$
\hti = (*)^{-1} \hi *,\qquad i=1,2,
$$
as in \pr{eqhht}. As before, the operators $\hi$ are formally
non-negative and are the formal adjoint of one another. If
$k=2,3,\dots,n$, let
$\o\in\Ll0k{\Htii}$ be a {\it zero mode} of
$\Htii$, i.e, a
$k$-form
$\o$ satisfying the equation $\Htii\o = 0$. In this section we
shall exclusively deal with formal eigenforms, so that in
particular $\o$ is {\it not}\/ required to be
square-integrable. By analogy with
\eq{mout}, it is natural to consider the equation
$$
d^+\,\bo = \delta^-\o
\Eq{bar}
$$
as an equation for the $(k-2)$-form $\bo$. The properties of
\eq{bar}, which are easily established, are analogous to those
of \eq{mout}, namely:

\item{i)} Equation \eq{bar} is compatible if and only if $\o$
is a zero mode of $\Htii$.
\item{ii)} $\bo$ is uniquely defined by \eq{bar} up to an
element of
$\Im d^+_{k-3}=e^\chi\Im d_{k-3}$. In particular, if $k=2$ we
have $\Im d_{-1}=\R$ (the space of constant functions on $M$),
so that $\Im d^+_{-1}=\R\,e^\chi$.
\item{iii)} Any solution $\bo$ of \eq{bar} is a zero mode of
$\Hti_{k-2}$, i.e., we have
$$
\Hti\bo = 0.
$$

From the above properties it follows that \eq{bar} defines a
mapping
$$
\overline{\phantom{H}}:\Ll0k{\Htii}/\Im\delta^-_{k+1}\to
\Ll0{k-2}{\Hti}/\Im d^+_{k-3},\qquad
2\le k\le n,
$$
which is easily seen to be an isomorphism by the assumption on
the de Rham cohomology of $M$. Note that for the above mapping
to be non-trivial
$M$ must be {\it non}\/-compact. Indeed, if $M$ is compact then
an easy integration by parts argument implies that
$\Ll0k{\Htii}=\Im\delta^-_{k+1}$ and
$\Ll0{k-2}{\Hti}=\Im d^+_{k-3}$.

Totally analogous considerations can be made for the components
$\Hii_k$ and
$\Hi_{k+2}$ of $H$. To be precise, consider the equation
$$
\delta^+\,\cho = d^-\o,
\Eq{check}
$$
where $\o$ is a $k$-form and $k=0,1,\dots,n-2$. As before,
the integrability condition for
\eq{check} is that $\o$ be a zero mode of $\Hii_k$, in which
case
$\cho\in\L^{k+2}(M)$ is uniquely defined modulo
$\Im\delta^+_{k+3}=e^\chi\Im\delta_{k+3}$ (with
$\Im\delta_{n+1}=*(\Im\delta_{-1})=\R\,\mu$), and is a zero
mode of
$\Hi_{k+2}$. Equation \eq{check} thus defines a mapping
$$
\breve{\phantom{H}}:\Ll0k{\Hii}/\Im d^-_{k-1}\to
\Ll0{k+2}{\Hi}/\Im \delta^+_{k+3},\qquad 0\le k\le n-2,
$$
which is again an isomorphism. Clearly, the mappings
$\overline{\phantom{\o}}$ and
$\breve{\phantom{\o}}$ are related by Hodge duality. More
precisely, an elementary calculation yields the following
result: 
\Pr{barcheck} The maps $(\breve{\phantom{\o}})_k$ and
$-(\overline{\phantom{\o}})_{n-k}$ are conjugated under Hodge
duality:
$$ (\breve{\phantom{\o}})_k = -(*)^{-1}\comp
(\overline{\phantom{\o}})_{n-k} \comp *.
\Eq{ovbr}
$$

Comparing equations \eq{mout} and \eq{bar}, we see that a
Moutard transform of a function
$\psi\in\Ll00{\Hii}\equiv\Ll00{H}$ on a two-dimensional manifold
$M$ is simply any function $\pt$ in the equivalence class of
$\overline{*\psi}$, which will be an element of
$\Ll00{\Hti}\equiv\Ll00{\Ht}$. This observation motivates the
following general definition:
\Df{moutard}  The {\it Moutard transform}\/ of
$\o\in\Ll0k{\Hii}/\Im d^-_{k-1}$ is the element
$$
\wo = \overline{*\o}
$$
of $\Ll0{n-k-2}{\Hti}/\Im d^+_{n-k-3}$.

The domain and the range of the {\it Moutard operator}\/
$\widetilde{\phantom{t}}$  defined above follow immediately
from the identity
$$
\widetilde{\phantom{\o}} = \overline{\phantom{\o}}\comp *,
$$
or equivalently, using \eq{ovbr},
$$
\widetilde{\phantom{\o}} = -\bigl(*\comp
\breve{\phantom{\o}}\bigr).
$$
In other words, if $\o\in\Ll0k{\Hii}/\Im d^-_{k-1}$ its Moutard
transform is the unique solution $\wo\in\L^{n-k-2}(M)/\Im
d^+_{n-k-3}$ of the equation
$$
d^+\wo = \delta^-(*\o),
\Eq{genmout}
$$
which is automatically an element of $\Ll0{n-k-2}{\Hti}/\Im d^+_{n-k-3}$.
As before, for the Moutard operator to be non-trivial
$M$ must be non-compact, since otherwise $\Ll0k{\Hii}=\Im
d^-_{k-1}$ and
$\Ll0{n-k-2}{\Hti}=\Im d^+_{n-k-3}$. Notice also that the
Moutard transform of a $k$-form $\o$ has different degree than
$\o$, unless $n=2m$ is even and
$k=m-1$, with $m=1,2,\dots$. For instance, when $m=1$ we have
$n=2$,
$k=n-k-2=0$, and we obtain the generalization of the classical
Moutard transformation to Riemannian surfaces introduced in the
previous section.
\Section{ex} Examples.
We present in this section a few examples of the
two-dimensional Darboux transformation on curved surfaces based
on the theory of quasi-exactly solvable Hamiltonians, \rf{GKO},
\rf{GKOqes}.

As our first example, consider the first-order differential
operators
$$
J^1 = \d_x\cq
J^2 = \d_y\cq
J^3 = x\,\d_x\cq
J^4 = x\, \d_y\cq
J^5 = y\, \d_y\cq
J^6 = x^2\,\d_x+x\, y\, \d_y  - 2\,x\,.
$$
The above differential operators span a Lie algebra
$\g\simeq\gl{2,\R}\semidirect\R^2$, which coincides with the
canonical form 1.11 in the classification of Lie algebras of
differential operators in two variables of reference
\rf{GKOreal} for $n = r=2$. The latter Lie algebra is {\it
quasi-exactly solvable}, \rf{GKO}, since it preserves the
subspace $\n\subset C^\infty(\R^2)$ whose elements are the
polynomials in the variables $(x,y)$ of total degree less than
or equal to 2. 

Let
$J$ denote the following element of the universal enveloping
algebra of
$\g$:
$$
\aeq{
J = (J^1)^2 &+ (J^2)^2 + 4\,(a-b)\,(J^3)^2
+ 4\,b\, (J^5)^2 + 4\,a\,(a-2b)\,(J^6)^2\\
&+ 2\,a\,\{J^3,J^5\}
-4\,(2a+b)\,J^3 + 2\,(a-12b)\,J^5+5(7b-a)\,,}
$$
where $a,b$ are real parameters such that $$a> 2b>0.\Eq{ab}$$
If we define
$$
\sigma = \frac14\log\bigl[1+2(a-2b)x^2\bigr]-
\frac32\log\bigl[1+2(a\,x^2+2\,b\,y^2)\bigr],
$$
then it can be shown that 
$$
-e^\sigma\cdot J\cdot e^{-\sigma} = h,
\Eq{hJ}
$$
where $h = -\Delta + V(x,y)$ is a Schr\"odinger operator on the
manifold $M=\R^2$ endowed with an appropriate metric. More
precisely (cf.~\rf{GKO}, Example 4.3.3), the contravariant
metric tensor
$(g^{ij})$ has components
$$
\aeq{
g^{11} &= (1+2\,a\,x^2)\bigl[1+2(a-2b)x^2\bigr],\\
g^{12} &= 2\,a\,x\,y\,\bigl[1 + 2(a - 2b)x^2\bigr],\\
g^{22} &= 1 + 4\,b\,y^2 + 4\,a\,(a - 2 b)\,x^2\,y^2,}
$$
and Gaussian curvature
$$
K = -2\,a\,\bigl[1 + 4(\,a - 2\,b)\,x^2)\bigr],
$$
and the potential $V(x,y)$ is given by
$$
V = -3\,a\,(a - 2\,b)\,x^2
- \frac b{1 + 2\,(a - 2\,b)\,x^2}
- \frac{48\,b}{1 + 2\,(a\,x^2 + 2\,b\,y^2)}.
$$
Since $J$ belongs to the universal enveloping algebra of $\g$ by
construction, it restricts to the finite-dimensional vector
space $\n$. We can therefore easily diagonalize $J|\n$,
obtaining the eigenvalues
$$\openup2\jot
\let\\\cr
\displaylines{
\l_0 = -5\,a + 23\,b + 4\,s,\qquad
\l_1 = -5\,a + 23\,b - 4\,s,\qquad
\l_2 = -a+3\,b,\\
\l_3 = -9\,a + 27\,b,\qquad
\l_4 = -3\,a + 15\,b,\qquad
\l_5 = -3\,a + 7\,b,
}
$$
where
$$
s = \sqrt{a^2 - 2\,a\,b + 9\,b^2}.
$$
Their corresponding eigenfunctions are given by
$$\openup2\jot
\aeq{
\ph_0 &= 3\,b + s + 2\,a\,(a - 2\,b)\,x^2,\qquad
\ph_1 = 3\,b - s + 2\,a\,(a - 2\,b)\,x^2,\\
\ph_2 &= 1 + 2\,a\,x^2 - 16\,b\,y^2,\qquad
\ph_3 = x,\qquad
\ph_4 = y,\qquad
\ph_5 = x\,y\,.}
$$
By \eq{hJ}, the Schr\"odinger operator $h$ possesses the six
eigenfunctions
$$
\psi_i = e^\sigma\,\ph_i,\qquad0\le i \le 5,
$$
with energies
$$
E_i = -\l_i.
\Eq{Ei}
$$
The first of these eigenfunctions has no zeros, and therefore it
must correspond to the ground state of $h$. The ground state
energy of $h$ is then given by
$$
E_0 = 5\,a - 23\,b - 4\,s,
$$
which is indeed manifestly lower than the remaining five
energies \eq{Ei} by
\eq{ab}.

To define the twisted Hamiltonian $H$ and the Darboux transform
of the eigenfunctions
$\psi_i$, it is convenient to take as $e^{-\chi}$ a constant
multiple of the 
ground state 
$\psi_0$, since this eigenfunction
has no zeros. We shall therefore define
$$
\chi = -\log(x^2 + c)-\sigma,
$$
where the constant $c>0$ is given by
$$
c = \frac{3\,b+s}{2\,a\,(a - 2\,b)}\,.
$$
According to \eq{h1}, the action of $H$ on a vector field $X$
can be expressed as
$$
(\hat H_1 X)^i = (-\nabla_j\nabla^j+V_1)\,X^i+
V^i{}_j\,X^j,
$$
where $V_1 = V-E_0+K$ and
$V^i{}_j=2\,\nabla^i\,\nabla_j\,\chi$. After a long but
straightforward calculation we find that
$$
V_1 = 
-7\,a + 23\,b + 4\,s - 11\,a\,\ga\,x^2 - \frac b{1 +
2\,\ga\,x^2} - 
  \frac{48\,b}{1 + 2\,(a\,x^2 + 2\,b\,y^2)},
$$
$$
\aeq{
V^1{}_2 &= 
-\frac{48\,a\,b\,x\,y\,(1 + 2\,\ga\,x^2)}{1 + 2\,(a\,x^2 +
2\,b\,y^2)}\,,\cr
V^2{}_1 &=
-\frac{48\,a\,x\,y\,(b + a\,\ga\,x^2 )}{1 + 2\,(a\,x^2 +
2\,b\,y^2)}
+ \frac{x\,y\,P_6(x)}{\ga\,a\,(c + x^2)^2\,(1 + 2\,\ga\,x^2)}
\,,}
$$
where
$$
\ga = a-2b>0
$$
and
$$
\aeq{
P_6(x) &=
48\,a^2\,\ga^3\,x^6+4\,a\,\ga^2\,(9\,a + 12\,b + 4\,s)\,x^4
+ 4\,\ga\,(5\,a^2 - 9\,a\,b - a\,s + 18\,b\,s+54\,b^2)\,x^2\cr
&\qquad{}+5\,a^2 - 22\,a\,b - 4\,a\,s + 30\,b\,s+90\,b^2\,.
}
$$
Note, in particular, that $V^1{}_2 \ne V^2{}_1$.

The Darboux transforms $d^-\psi_i$ of the eigenfunctions $\psi_i$,
$1\le i\le 5$, constructed above can also be computed in a
straightforward way using
\eq{dpsi}, with the following result:
$$
\aeq{
d^-\psi_1 &= \frac{4\,e^\sigma s\,x}{c+x^2}\,dx\\
d^-\psi_2 &=
2\,e^\sigma\,\left[\frac{x\,(16\,b\,y^2 + 2\,a\,c -
1)}{c+x^2}\,dx - 16\,b\,y\, dy\right]\\
d^-\psi_3 &=
e^\sigma\,\frac{c- x^2}{c+x^2}\,dx\\
d^-\psi_4 &=
e^\sigma\,\left[-\frac{2\,x\,y}{c+x^2}\,dx + dy\right]\\
d^-\psi_5 &= e^\sigma\,\left[\frac{y\,(c-x^2)}{c+x^2}\,dx +
x\,dy\right]\,. }
$$
As explained in \section{low}, the five vector fields
associated to these one-forms are formal eigenvectors of $\hat
H_1$ with eigenvalues $E_i-E_0$,
$1\le i\le5$. For example, the first of these vector fields is
$$
4\,s\,x\,\frac{1+2\,\ga\,x^2}{c+x^2}\,e^{\sigma}
\bigl[(1+2\,a\,x^2)\,\d_x
+  2\,a\,x\,y\,\d_y\bigr].
$$
In this case, it is easy to show that the five formal
eigenvectors of $\hat H_1$ constructed above are actually
square-integrable. Indeed, the square ${\rm L}^2$ norm of a
vector field
$X$ can be expressed as
$$
\norm X^2 = \int_{\R^2}\sqrt g\,g^{ij} X_i\,X_j\,dx\,dy,
$$
where $g = \det(g_{ij})$ and $X_i$ ($i=1,2$) are the components
of the associated one-form $X^\flat$. But
$$
e^{2\sigma} \sqrt g = (1 + 2\,a\,x^2 + 4\,b\,y^2)^{-7/2}<{\rm
const.}\,r^{-7},
\qquad r=\sqrt{x^2+y^2}\to\infty,
$$
and a straightforward calculation shows that
$e^{-2\sigma}g^{ij}X_i X_j$
is bounded by a constant times $r^4$ as $r\to\infty$ when
$X^\flat=d^-\psi_i$ for $i=1,\dots,5$, thus proving our
contention.

The limiting case
$$
a = 2b
$$
of the previous example is worth studying, because in this
case the curvature is constant and negative:
$$
K = -4\,b<0.
$$
Diagonalizing again $J|\n$ we find the following eigenvalues
$$
\l_0 = 25\,b,\qquad
\l_1 = 9\,b,\qquad
\l_2 = b,
$$
of respective multiplicities $1$, $2$ and $3$. The corresponding
eigenfunctions are
$$
\ph_0 = 1,\quad
\ph_1 = x,\quad
\ph_2 = y,\quad
\ph_3 = x\,y,\quad
\ph_4 = 12\,b\,x^2-1,\quad
\ph_5 = 12\,b\,y^2-1\,.
$$
As before, the six functions $\psi_i = e^\sigma\,\ph_i$ are
eigenfunctions of
$h$ with eigenvalue $E_i = -\l_i$, where $\sigma$ can be taken
as
$$
\sigma = -\frac32\log(1+4\,b\,r^2).
$$
Again, the first of these eigenfunctions has no zeros,
and therefore corresponds to the ground state, with ground energy
$$
E_0 = -25\,b.
$$
Taking $\psi_0 = e^{-\chi}$, or equivalently
$$
\chi = -\sigma\,.
$$
we find
$$
V_1 = 20\,b - \frac{48\,b}{1 + 4\,b\,r^2},\qquad
V^1{}_2 = V^2{}_1 = -\frac{96\,b^2\,x\,y}{1 + 4\,b\,r^2}\,.
$$
The Darboux transforms of the eigenfunctions $\psi_i$ ($1\le
i\le 5$) are easily computed. Their associated vector fields,
which are as usual eigenvectors of
$\hat H_1$ with eigenvalues $E_i-E_0$, are
$$
\openup2\jot
\Eqalign{
&e^\sigma\,\left[(1 +
4\,b\,x^2)\,\d_x+4\,b\,x\,y\,\d_y\right],
&&e^\sigma\,\left[4\,b\,x\,y\,\d_x+(1 +
4\,b\,y^2)\,\d_y\right],\cr 
&e^\sigma\,\left[(1 +
8\,b\,x^2)\,y\,\d_x+(1 + 8\,b\,y^2)\,x\,\d_y\right],
&&24\,b\,e^\sigma\,\left[(1 +
4\,b\,x^2)\,x\,\d_x+4\,b\,x^2\,y\,\d_y\right],
\cr
&24\,b\,e^\sigma\,\left[4\,b\,x\,y^2\,\d_x+ (1 +
4\,b\,y^2)\,y\,\d_y\right]\,.\cr
}
$$
Since
$$
e^{2\sigma}\,\sqrt g = (1+4\,b\,r^2)^{-7/2},
$$
an argument analogous to the one for the previous example
shows that the latter eigenvectors are all square integrable.

Consider now the canonical form 2.3 (with $n = 2$) in the
classification of reference \rf{GKOreal}, which is the Lie algebra
$\g\simeq\sL{3,\R}$ spanned by the first-order differential operators
$$
\openup2\jot
\let\\\cr
\displaylines{
J^1 = \d_x\,,\qquad
J^2 = \d_y\,,\qquad
J^3 = x\,\d_x\,,\qquad
J^4 = y\, \d_x\,,\qquad
J^5 = x\, \d_y\,,\qquad
J^6 = y\, \d_y\,,\\
J^7 = x^2\,\d_x+x\, y\, \d_y  - 2\,x\,,\qquad
J^8 = x\,y\,\d_x+ y^2\, \d_y  - 2\,y\,.}
$$
The latter Lie algebra leaves invariant the finite-dimensional
subspace
$\n\subset C^\infty(\R^2)$ introduced in the previous examples, so
that it is again quasi-exactly solvable. Let $J$ denote the element
of the universal enveloping algebra of $\g$ given by
$$
J = a\,\left[(J^1)^2 + \el\,(J^2)^2\right] + 
c\,\left[\el\,(J^7)^2 + (J^8)^2\right] 
+ b\,\left(\{J^1,J^7\}+\{J^2,J^8\}
+7\,J^3 + 7\,J^6\right)\,,
$$
where $a,c,\el$ are positive real
parameters, $b\ge0$, and
$$\ga = \el\,a\,c - b^2>0\,.\Eq{gamma}$$
If 
$$
\r = b + c\,(\el\,x^2+y^2)
\Eq{rho}
$$
and
$$
\sigma = -\frac58\,\log(\r^2+\ga) +
\frac74\,b\,\ga^{-1/2}\,\arctan(\ga^{-1/2}\,\r)\,,
$$
it is shown in \rf{GKO} that \eq{hJ} defines again a Schr\"odinger
operator $h$, with potential $V$ given by
$$
V = -\frac{71}4\,b - \frac34\,\r + \frac{(192\,b^2 -
45\,a\,c\,\el)\,\r +143\,a\,b\,c\,\el-192\,b^3}{4\,(\r^2+\ga)}\,.
$$
The metric of the Riemannian manifold $M=\R^2$ has now contravariant
components
$$
\aeq{
g^{11} &= a+x^2\,(b+\rho),\\
g^{12} &= x\,y\,(b+\rho),\\
g^{22} &= \el\,a+y^2\,(b+\rho),}
$$
and Gaussian curvature given by
$$
K = -2\,\r.
$$
The eigenvalues of $J|\n$  are easily found to be
$$
\l_0 = 4 \,b + 2\,s,\qquad
\l_1 =  4 \,b - 2\,s,\qquad
\l_2 = 12\,b\,,\qquad 
\l_3 = 2\,b\,,
$$
where
$$
s = \sqrt{16\,b^2 + 2\,a\,c\,\el}\,,
$$
and the last two eigenvalues have multiplicity 2.
The corresponding eigenfunctions are respectively
$$
\let\\\cr
\openup2\jot
\displaylines{
\ph_0 = \el\,x^2 + y^2 + \frac{s-4\,b}{c},\qquad
\ph_1 = \el\,x^2 + y^2 - \frac{s+4\,b}{c},\\
\ph_2 =  \el\,x^2 - y^2,\qquad
\ph_3 = x\,y,\qquad
\ph_4 = x,\qquad
\ph_5 = y\,.}
$$
On account of \eq{hJ}, multiplying these eigenfunctions of $J$ by the
factor $e^\sigma$ we obtain six eigenfunctions
$\psi_i=e^\sigma\,\ph_i$ of
$h$, with energies
$E_i = -\lambda_i$.
As in the previous examples, the first of these eigenfunctions
never vanishes, and therefore it corresponds to the ground state of
$h$, with ground energy given by
$$
E_0 = -4\,b-2\,s.
$$
We therefore take
$$
\chi = -\log\psi_0 = -\sigma - \log(\el\,x^2 + y^2 + k),
$$
with
$$
k = \frac{s-4\,b}{c} > 0\,,
$$
obtaining
$$
V_1 = 2\,s-\frac{55}4\,b - \frac{11}4\,\r + \frac{(192\,b^2 -
45\,a\,c\,\el)\,\r +143\,a\,b\,c\,\el-192\,b^3}{4\,(\r^2+\ga)}
$$
and
$$
V^1{}_2=\frac{2\,c\,x\,y\,P_3(\r)}{(\r^2+\ga)\,(\r +
s-5\,b)^2},\qquad V^2{}_1 = \el\, V^1{}_2,
$$
where
$$
\aeq{
P_3(\r) &= (27\,b -
4\,s)\,\r^3
+ (9\,a\,c\,\el + 14\,b\,s-79\,b^2)\,\r^2\\
&\hskip3em{}+ (317\,b^3 - 76\,b^2 s  -
16\,a\,b\,c\,\el + 6\,a\,c\,\el\,s)\,\r\\
&\hskip3em{}-201\,b^4 + 187\,a\,b^2 c\,\el -
50\,b\,\ga\,s + 14\,a^2\,c^2\,\el^2\,.
}
$$
Note that the denominator of $V^1{}_2$ never vanishes on account of
\eq{gamma}--\eq{rho}.

The Darboux transforms of the eigenfunctions $\psi_i$ ($1\le i\le
5$) are given by
$$
\openup2\jot
\aeq{
d^-\psi_1 &= \frac{4\,s\,e^\sigma}{c\,(k + \el\,x^2 +
y^2)}\,(\el\,x\,dx + y\,dy),\\
d^-\psi_2 &= \frac{2\,e^\sigma}{k + \el\,x^2 +
y^2}
\left[\el\,x\,(k + 2\,y^2)\,dx
- (k + 2\,\el\,x^2)\,y\,dy
\right],\\
d^-\psi_3 &= \frac{e^\sigma}{k + \el\,x^2 +
y^2}
\left[
y\,(k - \el\,x^2 + y^2)\,dx + x\,(k +\el\,x^2 - y^2)\,dy
\right],\\
d^-\psi_4 &= \frac{e^\sigma}{k + \el\,x^2 +
y^2}
\left[
(k - \el\,x^2 + y^2)\,dx-2\,x\,y\,dy
\right],\\
d^-\psi_5 &= \frac{e^\sigma}{k + \el\,x^2 +
y^2}
\left[
-2\,\el\,x\,y\,dx + (k + \el\,x^2 - y^2)\,dy
\right].
}
$$
The vector fields associated to these one-forms are eigenvectors of
$\hat H_1$ with eigenvalue $E_i-E_0$. As in the previous examples,
the latter vector fields are easily found to be square integrable on
$M$ on account of the asymptotic behavior at infinity of $e^{2\sigma}
\sqrt g$ and $e^{-\sigma}\,(d^-\psi_i)$.

\Section{ack}
Acknowledgments.
One of the authors (A.G.-L.) would like to thank A. Galindo, M.
Ma\~nas and M.~A.~Mart\'\i n-Delgado for helpful conversations.
\bigskip
\goodbreak
\References
\ends
\bye